\documentclass[twocolumn,twocolappendix]{aastex631}
\usepackage{multirow, hyperref}

\newcommand{\xmm}{{\em XMM-Newton}}

\newcommand{\fxu}{{erg s$^{-1}$ cm$^{-2}$}}

\newcommand{\lxu}{{erg s$^{-1}$}}

\received{September 11, 2025}
\accepted{November 4, 2025}

\shorttitle{Empirical Emission Measure distribution for solar type stars}
\shortauthors{A. Maggio et al.}
\graphicspath{{./}{figures/}}

\begin{document}

\title{Empirical prediction of plasma emission measure distributions and 
X-EUV spectra of late-type stars}

\correspondingauthor{Antonio Maggio}
\email{antonio.maggio@inaf.it}

\author[0000-0001-5154-6108]{A. Maggio}
\affiliation{INAF -- Osservatorio Astronomico di Palermo, Piazza del Parlamento, 1, I-90134, Palermo, Italy}

\author[0000-0003-4948-6550]{I. Pillitteri}
\affiliation{INAF -- Osservatorio Astronomico di Palermo, Piazza del Parlamento, 1, I-90134, Palermo, Italy}

\author[0000-0002-1600-7835]{J. Sanz-Forcada}
\affiliation{Centro de Astrobiolog\'{i}a (CSIC-INTA), ESAC Campus, E-28692 Villanueva de la Ca\~nada, Madrid, Spain}

\author[0000-0002-9900-4751]{G. Micela}
\affiliation{INAF -- Osservatorio Astronomico di Palermo, Piazza del Parlamento, 1, I-90134, Palermo, Italy}



\begin{abstract}
High-energy emission spectra from the outer atmospheres of late-type stars represent an important feature of the stellar activity in several contexts, such as the photoevaporation and photochemistry of planetary atmospheres or the modeling of irradiated circumstellar disks in young objects.
An accurate determination of these spectra in the EUV and soft X-ray (XUV) band requires high-resolution spectroscopy, that is rarely feasible with current instrumentation.
We employed a relatively large set of plasma emission measure distributions (EMDs) as a function of temperature, derived from FUV and X-ray emission line spectra acquired with the Hubble Space Telescope and Chandra or XMM-Newton, in order to devise a relatively simple recipe for predicting EMDs and XUV spectra of stars of different spectral types, activity levels, and plasma metallicity.
We show that the EMDs in the range of temperatures between 10$^4$\,K and 10$^{7.5}$\,K can be described using the stellar surface X-ray flux as a control parameter, but this parameterization also depends on the spectral type. In particular, we find that M-type stars show slightly lower emission measure at temperatures below $\sim 10^5$\,K and higher emission measures for $T \gtrsim 10^{6.5}$\,K with respect to FGK stars with similar surface X-ray fluxes.
We evaluated the uncertainties in the broad-band EUV and X-ray fluxes derived from synthetic EMDs and spectra, considering in the error budget also the limited knowledge of the chemical abundances in stellar outer atmospheres.  
\end{abstract}



\section{Introduction} \label{sec:intro}
High-energy emission in X-rays and UV bands (altogether XUV, $\sim 1$--920\,\AA) is a key feature that 
characterizes the evolution of the stars since their formation. 
XUV irradiation is responsible for the heating and evaporation of the protostellar disks, 
as well as the photochemistry of the surfaces of the disks and of the planetary atmospheres \citep{Ercolano2009, Owen+Alvarez2016, Picogna+2019, Locci+2024}. 

Observations with space-borne facilities, such as Hubble Space Telescope, Chandra or XMM-Newton, allow direct measurements of the stellar emission at FUV or X-ray wavelengths, at least in principle.
In practice, there are several difficulties. The limited sensitivity of current space instrumentation makes high-resolution spectroscopy feasible only for nearby or very active stars, that result sufficiently bright sources.
Moreover, severe interstellar absorption hampers a direct observation of the spectrum between the Lyman edge at 912\,\AA\ and shorter wavelengths in the soft X-ray band at about 100\,\AA, so that it was possible to obtain an EUV spectrum only for a handful of nearby stars and only up to $\sim 400$\,\AA\ \citep{Bowyer+2000,SanzForcada+2003}. 
The EUV spectral region has, however, a critical role related to the formation of the metastable He I line at 10.83\,$\mu$m, which constitutes a tracer of the atmospheric evaporation in highly irradiated planets 
\citep{SF25,Fossati2023,Poppenhaeger2022,Nortmann+2018}.

Computation of synthetic optically-thin spectra in the full XUV band requires a quite accurate reconstruction of the thermal structure of the outer atmosphere of solar-type stars, from the transition region up to the corona. The measurement of line fluxes of ions that form between $10^4$\,K and $10^7$\,K allows to determine the amount of plasma emitting in this temperature range, and to derive the so-called Emission Measure Distribution vs\, temperature (EM(T), or simply EMD). 
Again, this is possible by means of high-resolution spectra taken in far UV and X-rays, but with the same limitations as above.
Ideally, joint simultaneous observations in FUV and X-rays (e.g.\ with HST/COS and XMM-Newton/RGS as in \citealp{Maggio+2023,Maggio+2024}) are the best approach to this aim, because rapid time variability from flares can significantly change the EMD at high temperatures and hamper the analysis when using spectra taken at different epochs. 
However, coordinating different facilities is very difficult and the time available for such joint observation programs is often devoted to other science topics. 

The possible alternative to the optimal approach is the joint analysis of observations in the FUV, EUV, and soft X-ray bands at different epochs, neglecting possible variability.
This approach was employed several times in the past (e.g.\ \citealt{cno07,SF11,Bourrier+2020}), 
and it allowed to build empirical relationships between broad-band EUV and X-ray fluxes \citep{SF11, Johnstone+2021, SF25}.

The MUSCLES Treasury Survey \citep{France+2016,Youngblood+2016,Loyd+2016} addressed the issue of stellar irradiation in exo-planetary systems by
creating panchromatic (5\,\AA–5.5\,$\mu$m) spectral energy distributions
(SEDs) of M-type and K-type dwarfs, based on medium-resolution (CCD) X-ray spectra taken with XMM-Newton and Chandra, FUV spectra with Hubble Space Telescope,
empirical estimates of the
EUV emission \citep{Linsky+2014}, and stellar photospheric models. 
The survey was subsequently expanded to include additional M-type dwarfs and a few targets of F--G type (Mega-MUSCLES
survey, \citealt{Wilson+2021, Behr+2023}), but it remains focused mainly on low-mass low-activity stars. 

A more general approach was proposed by \cite{Wood+2018}, based on Chandra high-resolution spectroscopy of 19 main-sequence stars with different spectral types. They derived a grid of coronal EMDs as a function of the stellar surface X-ray flux, that can be used to estimate the distribution of the hot plasma for any main-sequence star with a measured broad-band X-ray luminosity. The main limitations of this work rest on the relatively small number of observed targets, mostly young/active G- and K-type stars, and the possibility to reconstruct EMDs only for temperatures $T \ge 10^{5.5}$\,K.

In this paper we present a novel empirical calibration of the stellar EMD, from chromosphere to corona, as a function of the surface X-ray flux for different spectral types. 
The present study enables the construction of synthetic EMDs starting from a very minimal information and the subsequent spectral synthesis especially in the EUV band.
We introduce our observational database in Sect.\ \ref{sec:data}, and a new parametric modeling in Sect.\ \ref{sec:emds}.
We discuss the results, the error budget, and limits of this approach in Sect.\ \ref{sec:discuss} and Sect.\ \ref{sec:ending}.

\begin{figure}
    \centering
    \resizebox{\columnwidth}{!}{
    \includegraphics[width=0.5\linewidth]{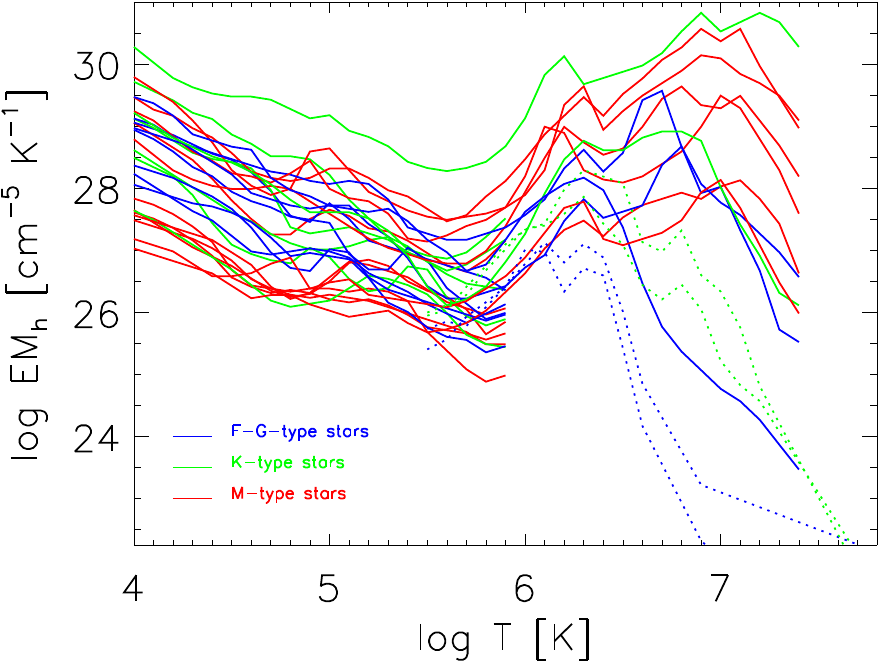}
    }
    \caption{Emission measure distributions of the sample stars. Truncated curves refer to Group\,1 (Table \ref{emduv}), while complete curves refer to Group\,2 (Table \ref{emdxuv}). Dashed lines indicate $\alpha$ Cen A and B (see text).
    }
    \label{fig:demd}
\end{figure}

\begin{figure*}
    \centering
    \includegraphics[width=0.8\linewidth,angle=0.]{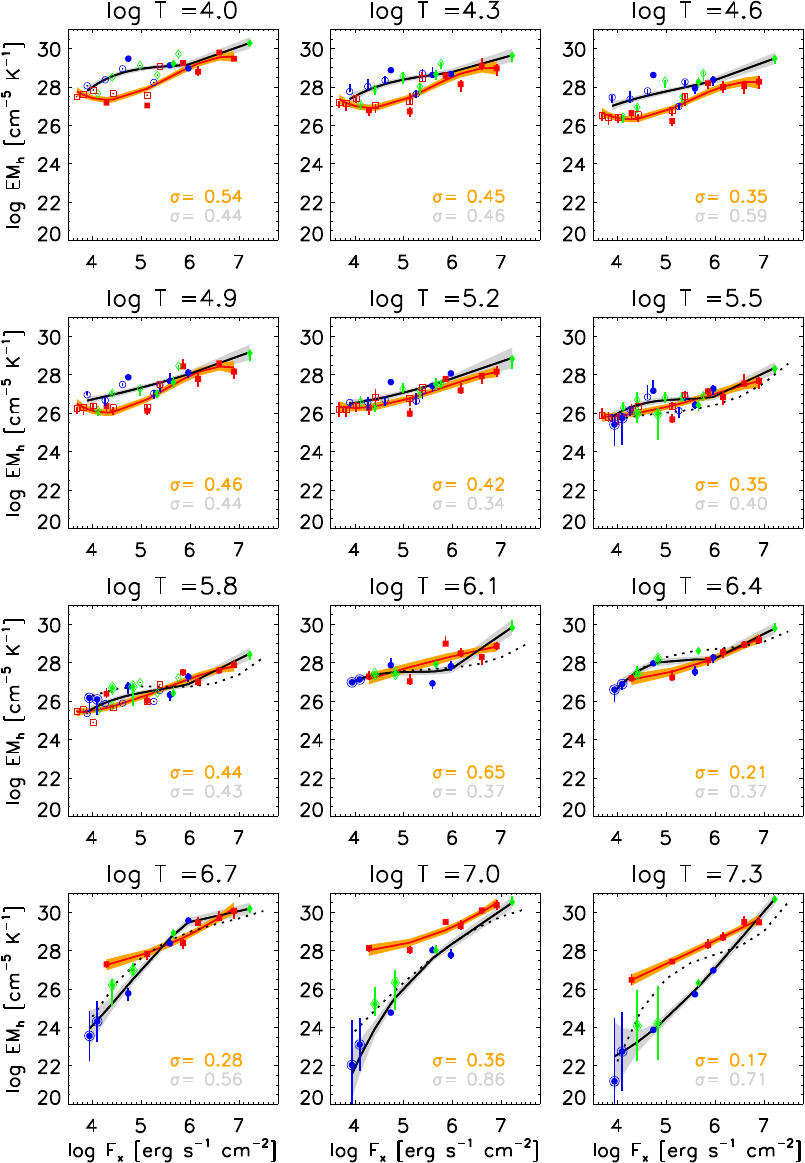}
    \caption{Differential Emission Measure at selected values of the plasma temperature vs.\ surface X-ray flux, for all stars in the sample. Blue symbols for F--G-type stars, green symbols for K-type stars, and red symbols for M-type stars. Empty and filled symbols identify stars in Group 1 or Group 2, respectively.
    Best-fit polynomials for FGK stars and M-type stars are shown with solid lines, the standard deviations are indicated on the bottom right of each panel, and shadowed bands represent $1\sigma$ confidence ranges. For temperatures $\log T \geq 5.5$, circumscribed symbols indicate $\alpha Cen$ A and B from \cite{Wood+2018}, and dashed lines show their best-fit polynomial results.}
    \label{fig:EMhFx}
\end{figure*}

\section{The EMD database}
\label{sec:data}
Our work is mainly based on the EMDs reconstructed from the analysis of high-resolution X-ray and FUV spectra, obtained from dedicated observation programs with spectrographs in several space missions, and published by \citet{SF25} (hereafter SF25). 
Four more stars were added to the main sample: Procyon \citep{SanzForcada+2003,SanzForcada+2004}, $\epsilon$\,Eri \citep{Chadney+2015}, $\iota$\,Hor \citep{SanzForcada+2019}, and V1298\,Tau \citep{Maggio+2023}.
The characteristics of these stars and their EMDs are reported in Appendix \ref{app:sample} (Tables \ref{emduv} and \ref{emdxuv}).

A large majority of the stars in our sample host exoplanetary systems, and they will be among the targets of detailed exoplanetary studies with the ESA mission Ariel \citep{Tinetti+2018}.

EMDs were reconstructed from measurements of the fluxes of a few tens of emission lines from ions of several chemical elements, with an homogeneous approach (cf. SF25). They are shown in Fig.\ \ref{fig:demd}. 

The original volume emission measures are defined as:
\begin{equation}
    EM_{\rm v}(T) = n_{\rm e} n_{\rm H} \Delta V  
\end{equation}
where $n_{\rm e} n_{\rm H}$ is the product of the electron and proton densities, and they are derived in a number of temperature bins $\Delta \log T = 0.1$. At the same time, the analysis of high-resolution FUV/X-ray spectra provides measurements of the chemical abundances of the line-emitting elements.

Our sample of EMDs is composed of two main groups.
Group 1 includes 16 stars for which we have only FUV high resolution spectra. 
These are stars with low X-ray fluxes and for which it is difficult to acquire a high resolution spectrum in X-ray within reasonable exposure times.
The FUV spectra are dominated by emission lines of elements in low-ionization states, and hence the corresponding EMDs are derived in the temperature\footnote{Hereafter, $T$ is in K units.} range $\log T = 4.0-5.9$ (20 temperature bins in chromosphere and transition region; see Table \ref{emduv}).

Group 2 is composed by 11 stars bright in X-rays for which we have both FUV and X-ray high resolution spectra. For these stars full EMDs are available in the temperature range from $\log T = 4.0$ to $\log T = 7.4$ (35 temperatures bins). 
These EMDs allow to probe the thermal structure of the outer stellar layers from the chromosphere to the corona (Table \ref{emdxuv}).

We complemented our study with four EMDs obtained from the analysis of Chandra high-resolution spectra, as reported in \cite{Wood+2018}, corresponding to the low-activity and high-activity states of $\alpha$\,Cen A (G-type) and $\alpha$ Cen B (K-type). 
This addition provides us with examples of coronal EMDs (20 bins in the range $\log T = 5.5$--7.4) for low-activity solar-type stars, down to a Sun-like regime.

We scaled the EMDs by the area of the stellar hemisphere, and computed the differential emission measure distributions (Fig.\ \ref{fig:demd})
\begin{equation}
    EM_{\rm h}(T) = n_{\rm e} n_{\rm H} \frac{\Delta h}{\Delta \log T} = \frac{EM_{\rm v}(T)}{2 \pi R_*^2 \Delta \log T}
\end{equation}
following the parametrization used by \cite{Wood+2018}.

The general shape of these EMDs can be described with a decreasing amount of emitting plasma down to $\log T = 5.5$--5.9, an increasing $EM_{\rm h}(T)$ for higher temperatures (in corona), and a variety of tails at the highest temperatures, with peaks occurring in the range $\log T = 6.5$--7.1, depending on the spectral type and the average activity level of the stars.

Observations with XMM-Newton of stars in Group 1 and Group 2 provided also medium-resolution (CCDs) X-ray spectra, that are feasible even for low-activity stars with photon counting rates not sufficient for high-resolution spectroscopy. 
Assuming simple coronal models with a low number (1--3) of isothermal components, previous analyses of medium-resolution X-ray spectra (see references in Appendix \ref{app:sample}), supplied also the effective temperature and emission measure of each discrete component, the plasma metallicity, and broad-band X-ray fluxes (0.1--2.4\,keV or 5--100\,\AA).
For the stars in Group\,1, this information combined with the $EM_{\rm v}(T)$ at low temperatures derived from high-resolution FUV spectra, yielded "hybrid" EMDs that allowed computing synthetic spectra and deriving also broad-band (100--920\,\AA) EUV fluxes (SF25).

Finally, for all the stars in our sample we gathered stellar distances (from GAIA DR3 parallaxes) and stellar radii (see Tables \ref{emduv} and \ref{emdxuv} and references therein), in order to compute surface X-ray fluxes
\begin{equation}
    F_{\rm x} = f_{\rm x} \frac{D^2}{R_*^2}
\end{equation}
where $f_{\rm x}$ is the stellar flux at Earth corrected for interstellar absorption.

Our work targets especially those XUV faint stars for which broad-band X-ray fluxes can be measured from low-resolution CCD spectra, but high-resolution X-ray or FUV spectra cannot be obtained in reasonable exposure times with current instrumentation.
This situation motivated the adoption of the stellar surface flux $F_{\rm x}$ that can be obtained  from 1-T or 2-T fits of CCD-resolution X-ray spectra as the main control parameter for computing synthetic EMDs.


\begin{figure*}
    \centering
    \includegraphics[width=0.45\linewidth,angle=180.]{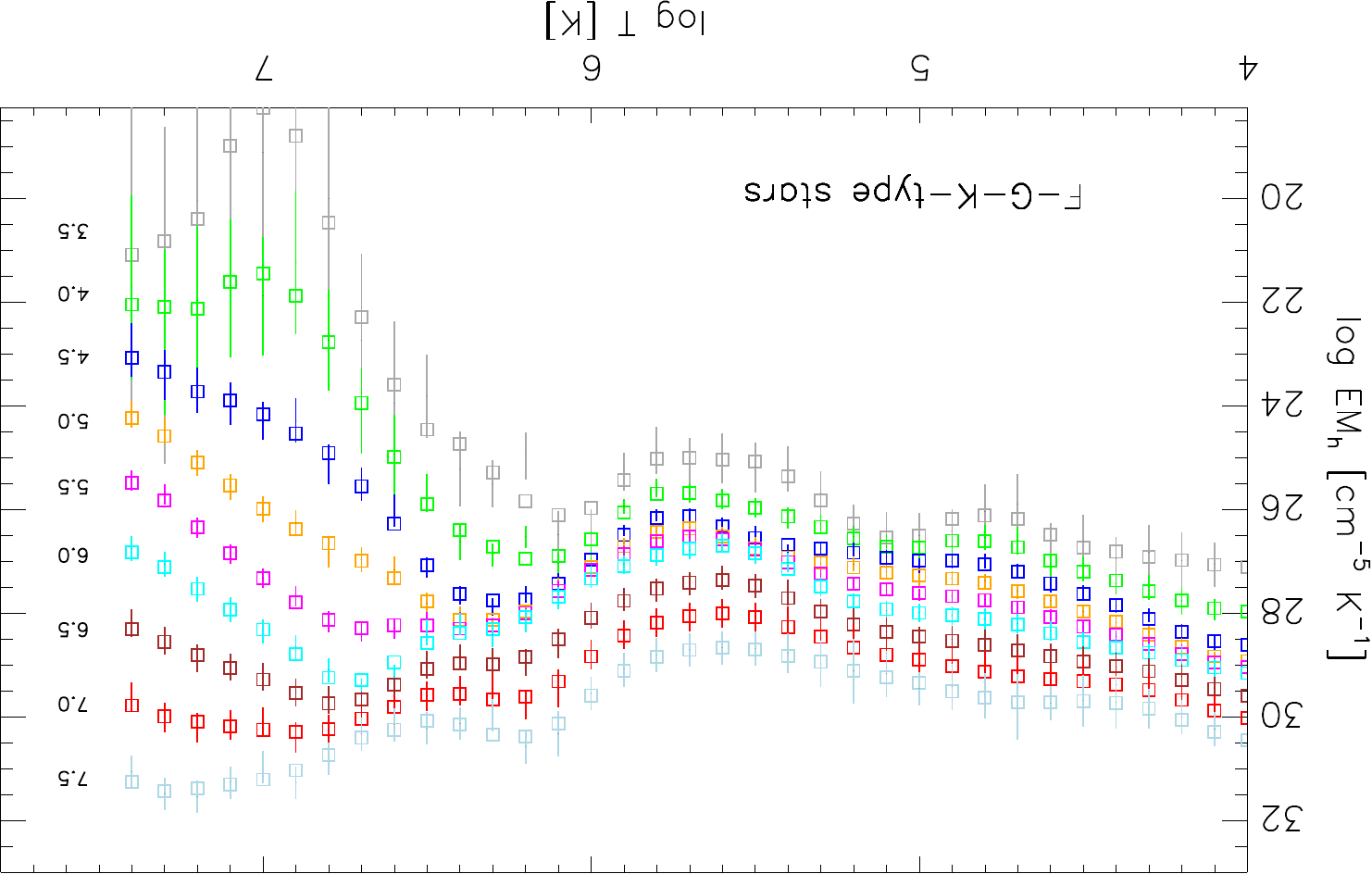}
    \includegraphics[width=0.45\linewidth,angle=180.]{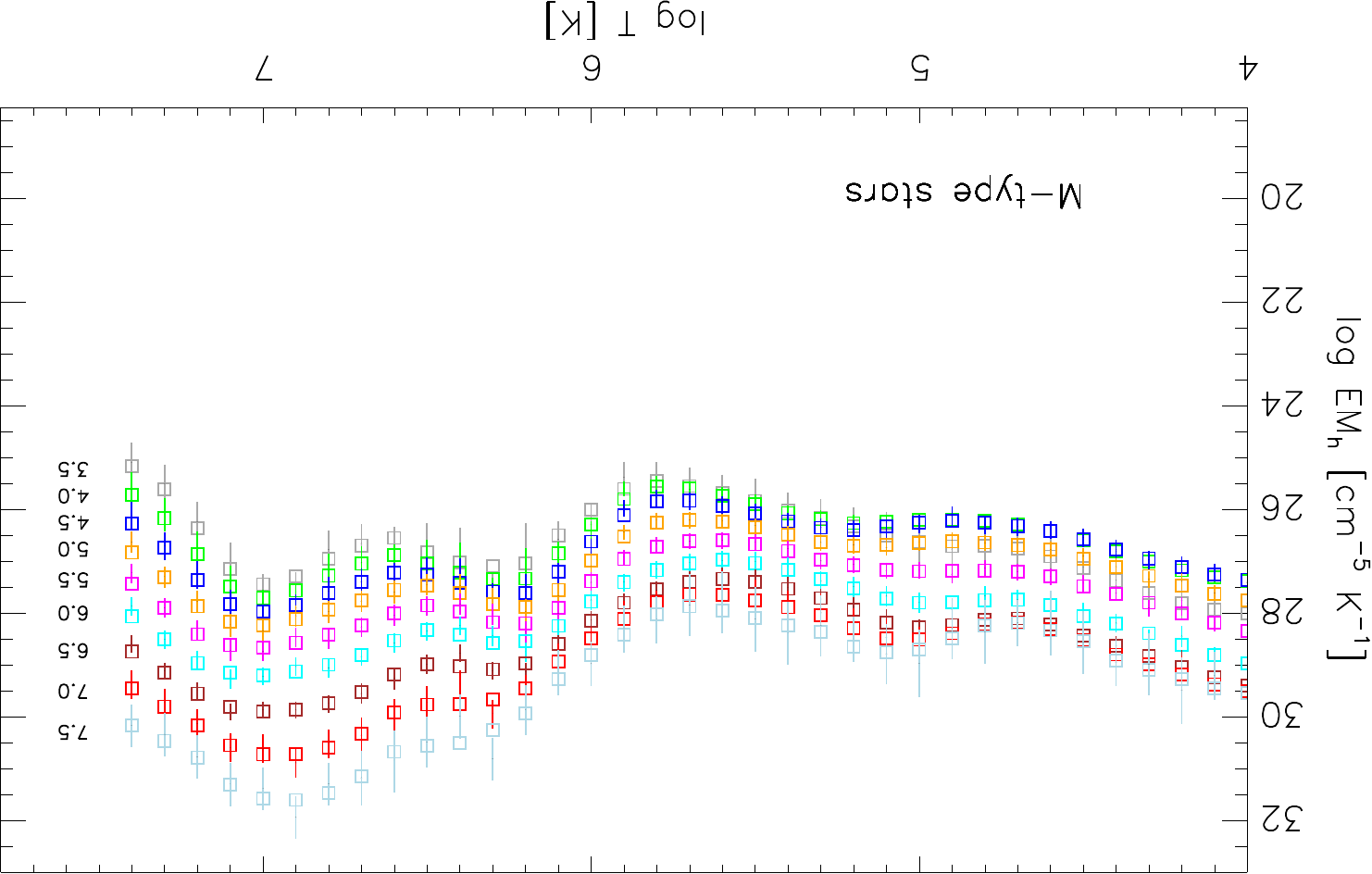}
    \caption{Synthetic emission measure distributions at selected reference values of the stellar surface X-ray flux (values on a log-scale indicated on the right side of each panel). The $EM_{\rm h}$ grid for FGK-type stars is on the left panel, and on the right panel for M-type stars. A 6th degree polynomial smoothing was applied, and 1$\sigma$ error bars are shown.}
    \label{fig:grids}
\end{figure*}

\begin{figure}
    \centering
    \includegraphics[width=\linewidth]{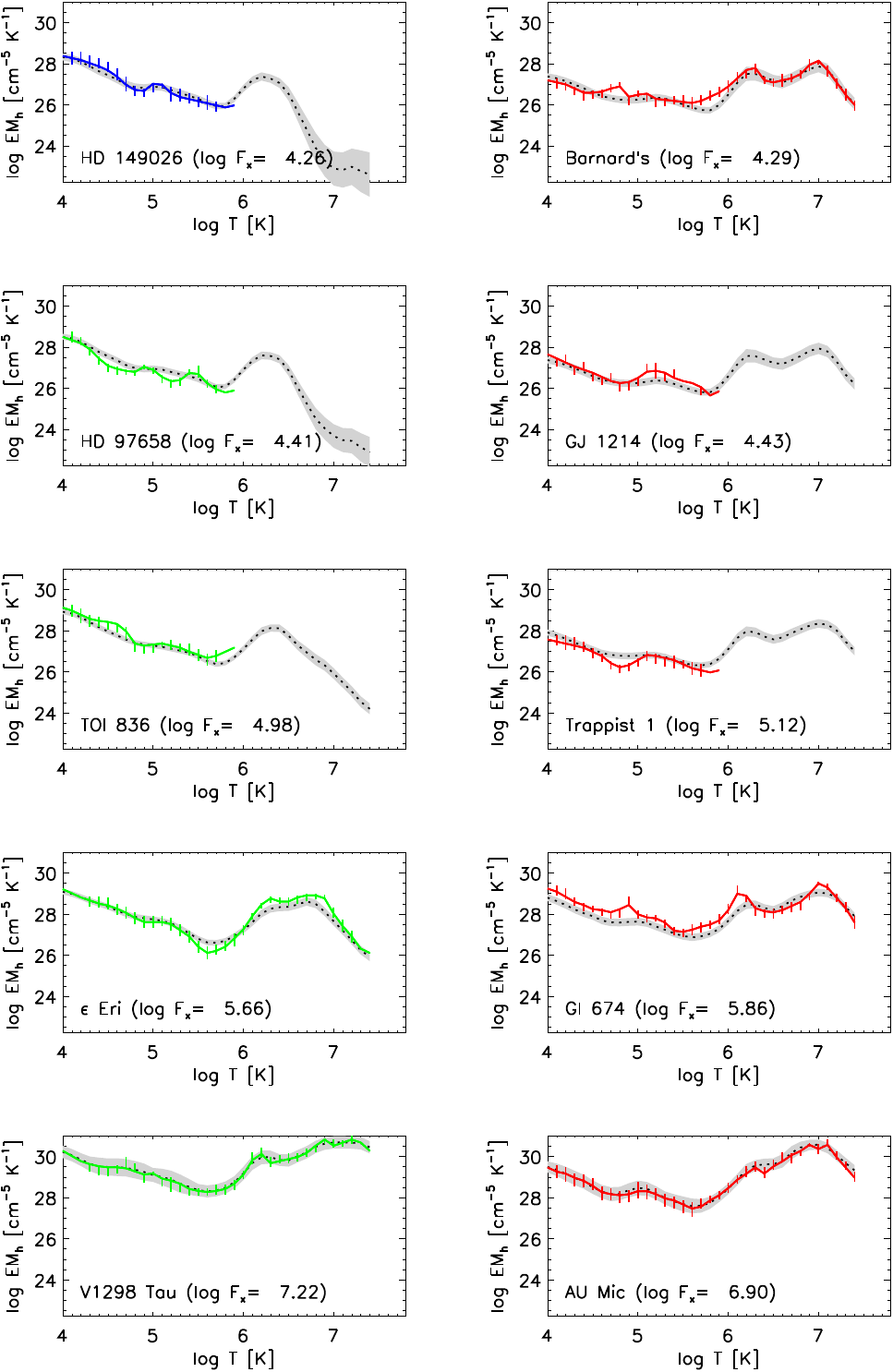}
    \caption{Reconstructed and synthetic emission measure distributions for selected stars in the original sample. G--K-type stars (on the left) are compared with M-type stars (on the right) having similar surface X-ray flux. The color coding of the original EMDs (solid lines) is the same as in Fig.\ \ref{fig:demd}. Dotted lines and gray-shaded $1 \sigma$ uncertainty regions were obtained by interpolating the curves in Fig.\ \ref{fig:grids}.
    }
    \label{fig:compare}
\end{figure}

\section{Parameterization of EM(T,$F_{\rm x}$)}
\label{sec:emds} 

Following \cite{Wood+2018}, we considered the values of $EM_{\rm h}(T_{\rm i})$ vs.\ $F_{\rm x}$, 
in each bin of temperature $T_{\rm i}$, for all the sample stars (Fig.\ \ref{fig:EMhFx}).
We realized that there are systematic differences in the trends for FGK stars (19 EMDs, 10 from 
Group 1, 5 from Group 2, 4 relative to $\alpha$ Cen A and B) with respect to those of M-type stars 
(12 EMDs, 6 from Group 1, and 6 from Group 2). Hence, we decided to treat separately these two different spectral type subgroups.

We computed best-fit polynomials to approximate the observed trends. We employed polynomials of degree 3, except for the subsample of M stars in the temperature bins $\log T > 5.9$, where we can rely on only six EMDs. 
In this case, polynomials of degree 2 were sufficient to describe the $EM_{\rm h}$ vs.\ $F_{\rm x}$ functional relation.
For comparison, in Fig.\ \ref{fig:EMhFx} we also show the third order polynomials used by \cite{Wood+2018} to fit the  $EM_{\rm h}$, but valid only for temperatures $T \geq 10^{5.5}$\,K, and determined upon stars of any spectral type. 

We remark that the four subsamples of EMDs cover slightly different ranges of 
stellar X-ray surface fluxes:
FGK stars in Group 1 cover the range $F_{\rm x} = 10^{3.9}$--$10^{5.8}$\,erg\,s$^{-1}$\,cm$^{-2}$, while those in Group 2 cover the range $10^{4.7}$--$10^{7.2}$\,erg\,s$^{-1}$\,cm$^{-2}$. 
With the addition of $\alpha$ Cen A, we extend also the range for G-type stars with available coronal EMDs down to $F_{\rm x} = 10^{3.9}$\,erg\,s$^{-1}$\,cm$^{-2}$. 
The M-type stars span the $F_{\rm x}$ ranges $10^{3.7}$--$10^{5.4}$\,erg\,s$^{-1}$\,cm$^{-2}$ in Group 1, and $10^{4.3}$--$10^{6.9}$\,erg\,s$^{-1}$\,cm$^{-2}$ in Group 2. This is relevant for the limits of validity of our polynomial fits.

We used the above results to build a synthetic grid of EMDs for a reference set of $\log F_{\rm x,ref}$ values, ranging from 3.5 to 7.5, in steps of 0.5 dex. We performed a 6th-degree polynomial smoothing, in order to remove some numerical noise. The result is shown in Fig.\ \ref{fig:grids} and tabulated in Appendix \ref{app:recipe} (Tables \ref{tab:fgksynthgrid} and \ref{tab:msynthgrid}).
Note that the extreme values of the $\log F_{\rm x,ref}$ range are slightly outside the ranges covered by the original stellar sample.
In order to avoid runaway $EM_{\rm h}(T)$ values, we employed a linear extrapolation, rather than using the best-fit polynomials.
Nonetheless, the synthetic EMDs for stars at these extremely low or high $F_{\rm x}$ values remain affected by relatively larger uncertainties.

\section{Results and discussion} 
\label{sec:discuss}
The main outcome of our study is the computation and parametrization of synthetic EMDs as a function of surface X-ray fluxes and temperatures for FGK and M-type stars, respectively (Sect.\ \ref{sec:emds}).

\begin{figure*}
    \centering
    \includegraphics[width=0.95\linewidth]{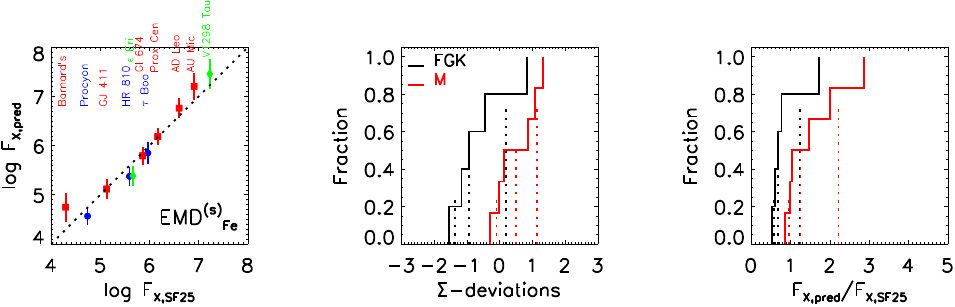}
    \includegraphics[width=0.95\linewidth]{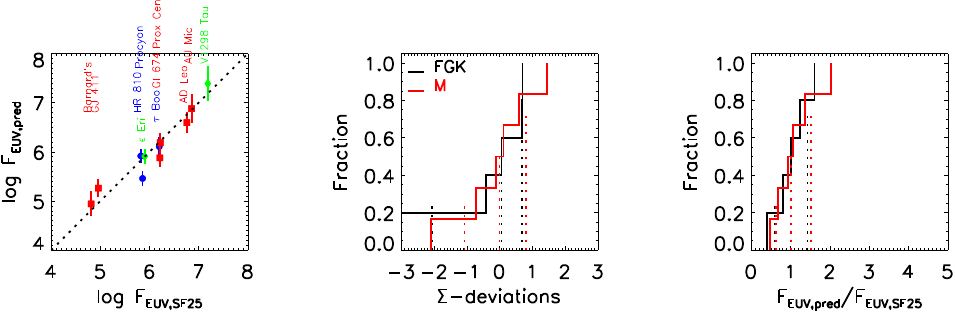}
    \caption{Comparison of measured X-ray and EUV surface fluxes with values based on synthetic EMDs. This is the case for stars in Group 2 and iron abundance fixed to the best-fit values for each star. Top row: scatter plot of X-ray fluxes (left), and cumulative distributions of $\Sigma$--deviations (center) and flux ratios (right). Vertical dotted lines mark the median values and the 25\% and 75\% quantiles. Bottom row: similar plots, but for the EUV surface fluxes.
    }
    \label{fig:errors_Fe}
\end{figure*}

\begin{figure}
    \centering
    \includegraphics[width=0.494\linewidth]{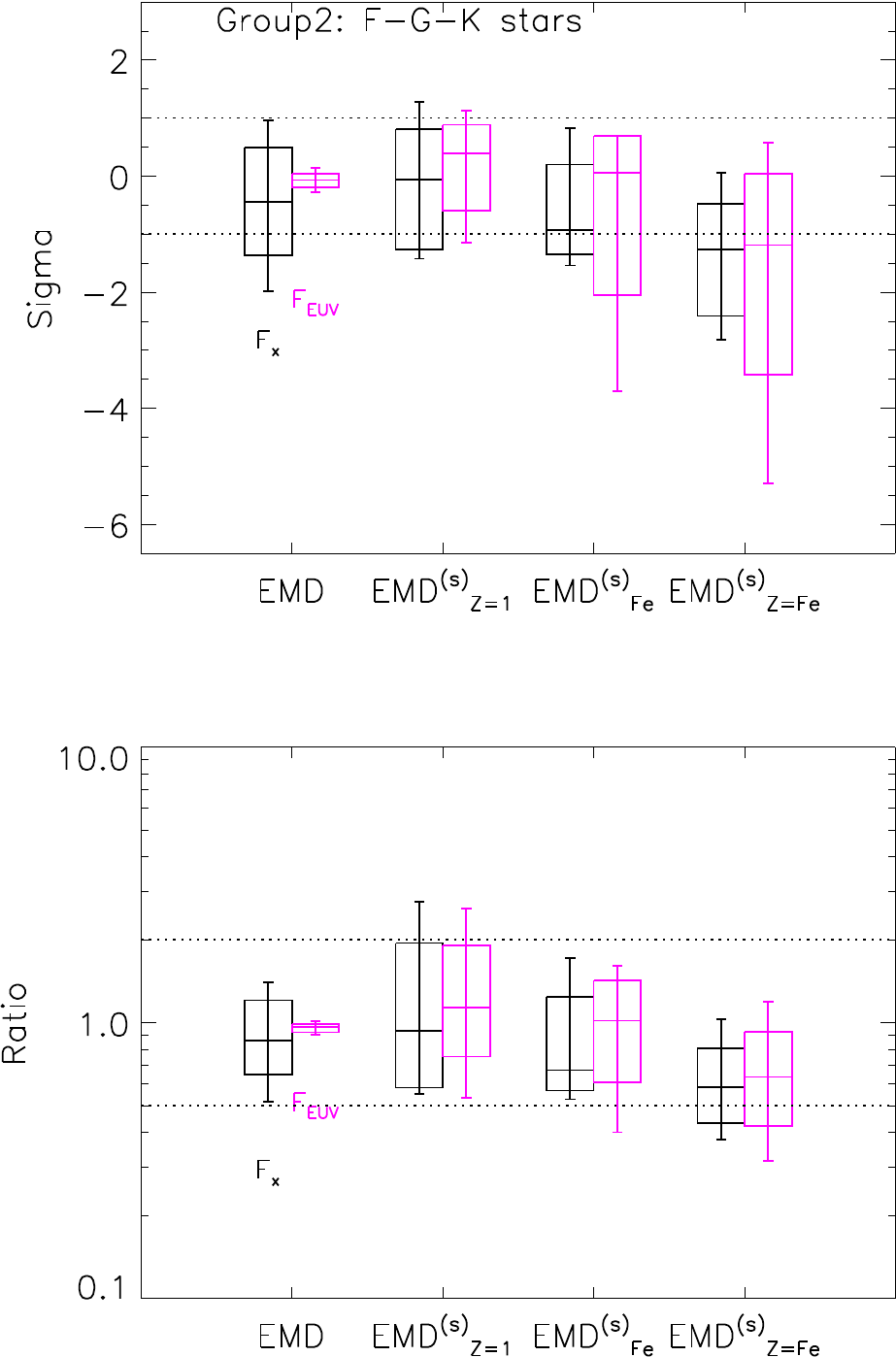}
    \includegraphics[width=0.494\linewidth]{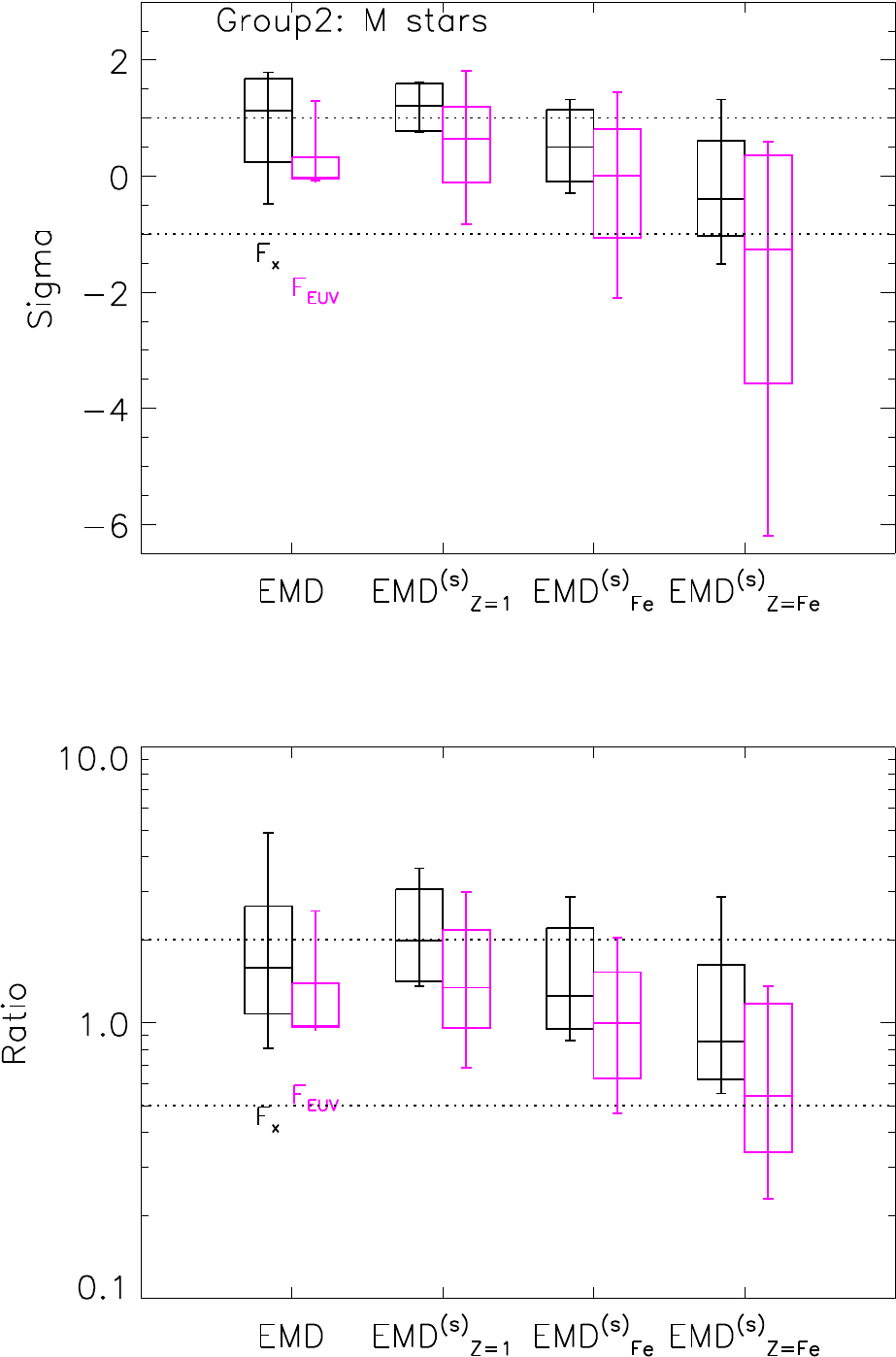}
    \caption{Comparison between measured and predicted (synthetic model) stellar surface fluxes for stars in Group 2. Top row: boxplots of the distributions of the $\sigma$-deviations between measured and computed X-ray fluxes (in black) or EUV fluxes (magenta), for stars with different spectral types (see legend). The central line in each box indicate the median, the bottom and top sides correspond to the central 50\% of the distributions, while the vertical lines span the full range of values. The labels on the X-axis represent the four cases discussed in the text.
    The dotted lines represent the $\pm 1\sigma$ uncertainty band. Bottom row: corresponding boxplots of the distributions of $F_{\rm x, pred}/F_{\rm x, obs}$ ratios. The dotted lines represent a factor 2 with respect to $F_{\rm x, obs}$, in opposite directions.
    }
    \label{fig:boxplots2}
\end{figure}

At temperatures below $T \sim 10^5$\,K, the M-type stars show lower $EM_{\rm h}$ than FGK stars, while at $T > 10^{6.5}$\,K, the $EM_{\rm h}$ in the coronae of M-type stars are systematically higher than those of earlier spectral types (panels in the top row of Fig.\ \ref{fig:EMhFx}).
At high temperatures the divergence between the $EM_{\rm h}$ of FGK and M-type stars is more marked when the 
surface flux is lower (panels in the bottom row of Fig. \ref{fig:EMhFx}).
In the intermediate temperature range, the trends of $EM_{\rm h}$ vs.\ $F_{\rm x}$ are similar.

In other words, the FGK stars with relatively low $\log F_{\rm x} < 5.5$ tend to have values of the emission measure in chromosphere and transition region higher than the M-type stars, while the low-$F_{\rm x}$ M-type stars show significantly more high-temperature plasma than the FGK stars with similar emission levels.

Overall, we found deviations of the reconstructed emission measure, $EM_{\rm h}(T)$, with respect to the best-fit polynomial approximation always $\le 3\sigma$\footnote{$\sigma$ here is the uncertainty on individual $EM_{\rm h}(T)$ values}
in any temperature bin
(Fig.\ \ref{fig:compare}). 
Few stars show relatively large deviations (see Fig. \ref{fig:compare_more} in Appendix \ref{app:addfigs}), in particular WASP\,13 at $\log F_{\rm x} = 5.26$ is a G-type star with chromospheric $EM_{\rm h}$ depressed with respect to other stars with similar spectral type, and more in line with M-type stars; on the contrary, Procyon ($\log F_{\rm x} = 4.73$) 
exhibits excess emission measure in chromosphere with respect to other FGK stars. 
Two more M-type stars show noticeable deviations with respect to the mean trend: 
GJ\,3470 at $\log F_{\rm x} = 5.38$ is an M-type star with a behavior more similar to stars with earlier spectral types, while GJ\,411 ($\log F_{\rm x} = 5.13$) rests below the polynomial approximation valid for other M-type stars.

In most of the coronal temperature bins where the differences related to the spectral type are small, we found a good agreement with the results by \cite{Wood+2018}. In the highest temperature bins, the \cite{Wood+2018} solutions represent better the behavior of the FGK stars rather than the M-type stars, that deviate significantly. In fact, their sample includes only two M-type stars, AD\,Leo and AU\,Mic, i.e. those with the highest surface X-ray fluxes in our sample.

In order to test the validity and accuracy of our synthetic EMDs, beyond a direct comparison with the original EMDs (Fig. \ref{fig:compare}), we computed synthetic spectra\footnote{Plasma emissivities derived from the ATOMDB v3.0.9 database \citep{aped,Foster+2020}.} in the wavelength range 1.24--1240\,\AA\ and broad-band surface X-ray and EUV fluxes in the ranges 5--100\,\AA and 100--920\,\AA, respectively. Then we performed a comparison of these predicted fluxes, $F_{\rm x,pred}$ and $F_{\rm euv,pred}$, with the original values $F_{\rm x,SF25}$ and $F_{\rm euv,SF25}$ (\citealt{SF25}, Table C.3).

We evaluated the differences in a statistical way, by computing the cumulative distributions of the $\sigma$-deviations, $\Sigma = (F_{\rm x, pred}-F_{\rm x, SF25})/(\sigma_{\rm x, SF25})$, between the measured (best-fit $F_{\rm x, SF25}$) and predicted ($F_{\rm x, pred}$, based on EMDs) surface X-ray fluxes. We also computed the corresponding distributions of the flux ratios $R = F_{\rm x, pred}/F_{\rm x, SF25}$. 
In a similar way, we also compared the measured and predicted EUV fluxes.
We show the result of such a comparison in Figure \ref{fig:errors_Fe}, for just one specific case (case {\it c}, described below).

\begin{figure}
    \centering
    \includegraphics[width=0.494\linewidth]{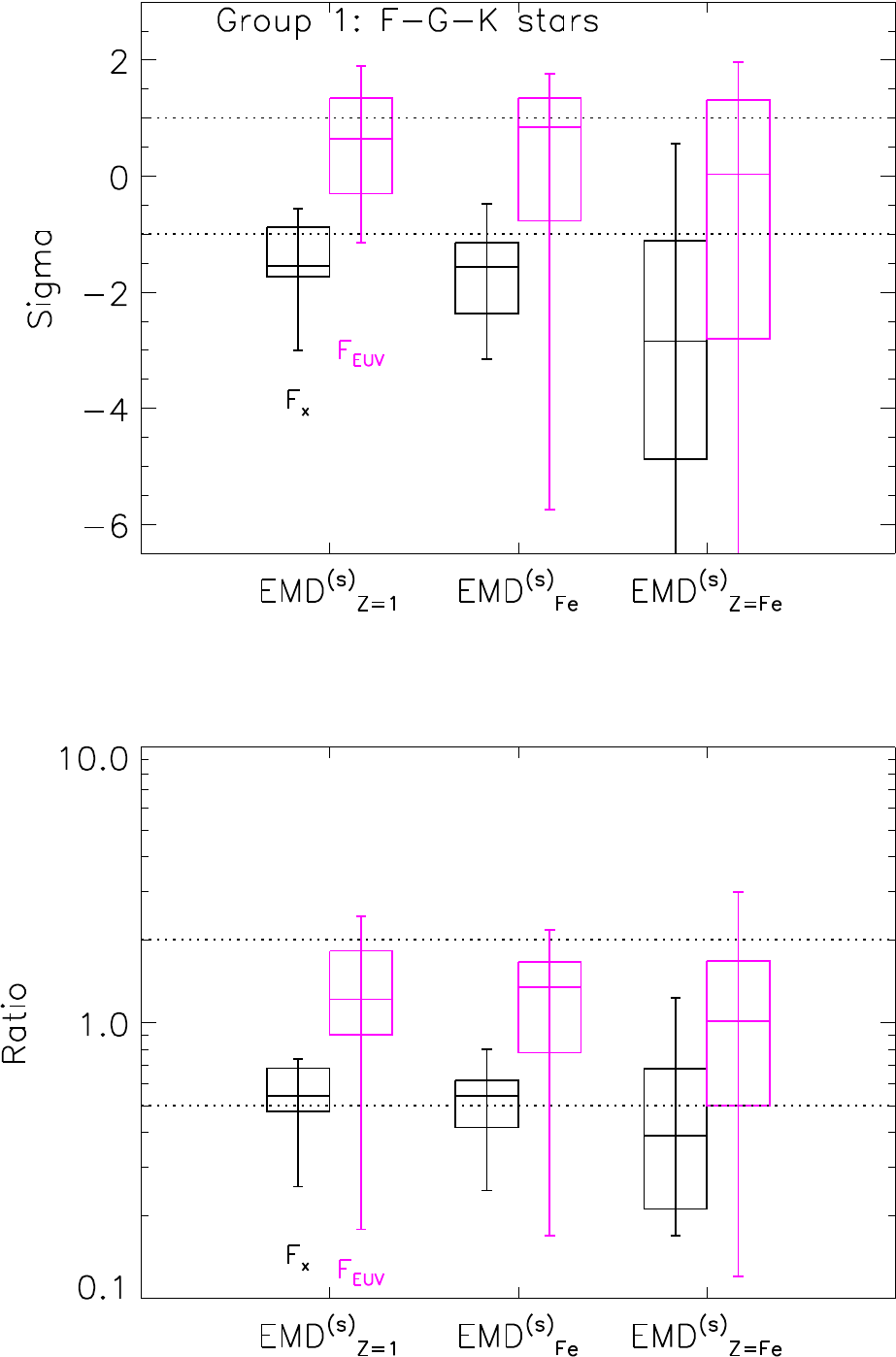}
    \includegraphics[width=0.494\linewidth]{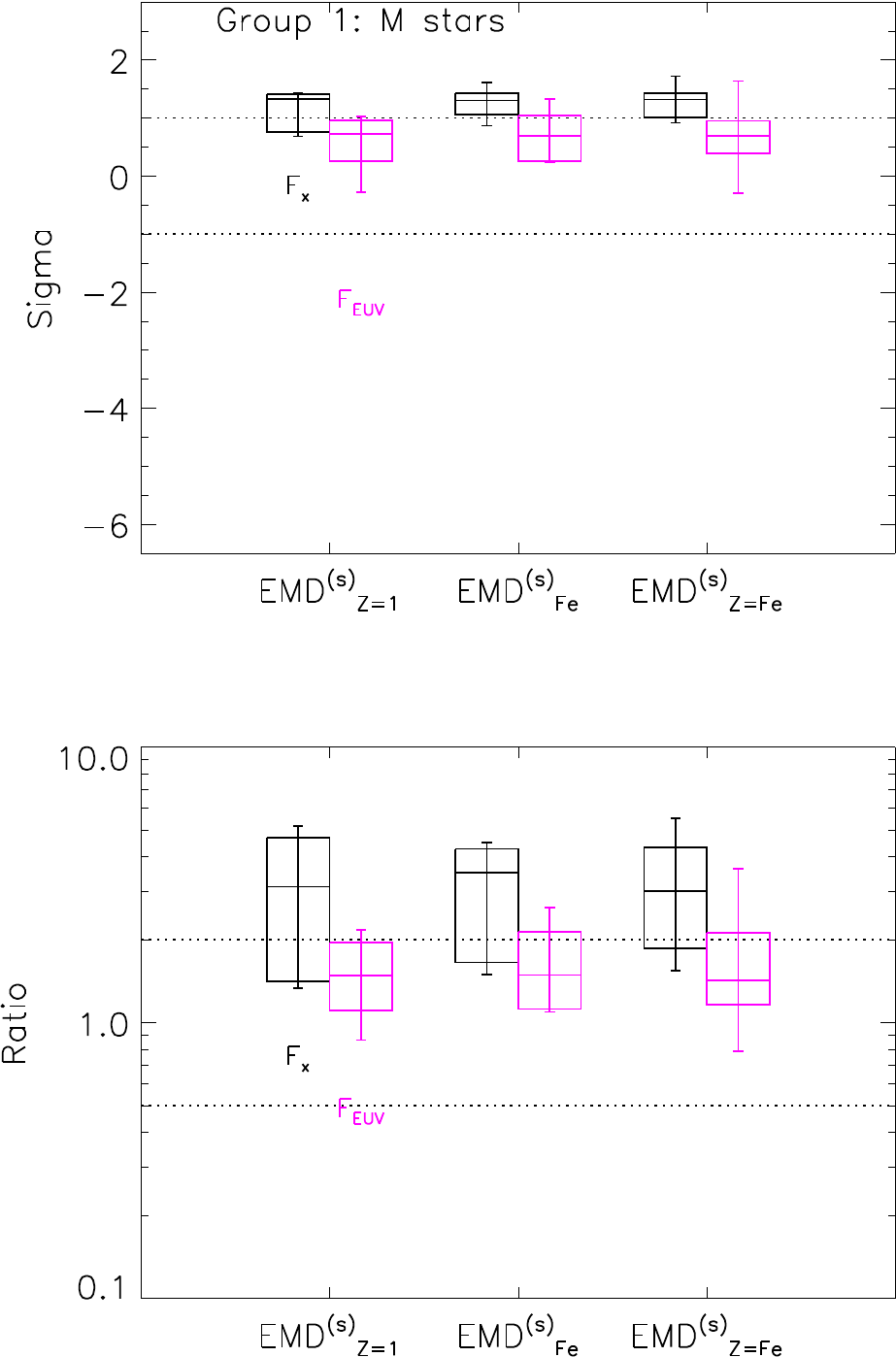}
    \caption{Comparison between measured and predicted (synthetic model) stellar surface fluxes for stars in Group 1. Boxplots in the top and bottom rows are similar to those in Fig.\ \ref{fig:boxplots2}, except that only three cases are considered.
    }
    \label{fig:boxplots1}
\end{figure}

There are several possible sources of uncertainty in predicting XUV spectra and hence X-ray or EUV fluxes, due to the analytic modeling of the synthetic EMDs, and our limited capability to fix the plasma metallicity or individual element abundances, that enter as weighting factors of the plasma emissivity.
We evaluated the systematic error due to these effects by computing $F_{\rm x, pred}$ and $F_{\rm euv, pred}$ with different assumptions. 
We summarize them in the following and compare schematically the results with boxplots of the $\Sigma$ and $R$ distributions, as shown in Fig.\ \ref{fig:boxplots2} and Fig.\ \ref{fig:boxplots1}.

\begin{center}
{\it (a) Original EMD case}
\end{center}
This is a reference case, in which $F_{\rm x, pred}$ are computed from the original EMDs and associated chemical abundances. The comparison with the $F_{\rm x,SF25}$ values, adopted as control parameter for the synthetic EMDs, is useful to quantify the difference between measurements based on full EMDs and the results of fitting CCD-resolution spectra. This is a sort of basal component of the error budget, that affects also the fluxes derived in the following three cases.

\begin{center}
{\it (b) Solar abundances (EMD$^{\rm (s)}_{Z=1}$ case)}
\end{center}
In this case and in the following we employed the synthetic EMDs. Here we assumed all chemical abundances fixed to solar values. This case represents the lowest level of knowledge about the true coronal plasma abundances.

\begin{center}
{\it (c) Fixing iron abundance only (EMD$^{\rm (s)}_{\rm Fe}$ case)}
\end{center}
The iron abundance is assumed equal to the value obtained from the spectral fit with multi-component isothermal models, while all other elements remain fixed to solar abundances. 
This is the most realistic scenario in the application of our methodology, due to the lack of knowledge about the other coronal abundances.
We selected the results of this case for the example shown in Figure \ref{fig:errors_Fe}.

\begin{center}
{\it (d) Scaling all abundances to iron (EMD$^{\rm (s)}_{Z=Fe}$ case)}
\end{center}
A final possibility is to assume the abundance of all elements equal to the measured iron abundance. 
This assumption introduces a bias for stars where the coronal abundance of elements with low First Ionization Potential are higher or lower than in photosphere (solar-like vs.\ inverse FIP effect, \citealt{Laming+2015}).
In our sample, about half of the stars in both Group 1 and Group 2 have subsolar iron abundances, that we attribute to an inverse FIP effect.
Consequently, if all abundances are fixed to the low iron value the resulting synthetic spectrum yields underestimated X-ray and EUV fluxes.

In the following we describe the main characteristics of the results in the cases (b) and (c) above. More details on our evaluation of the error budget 
are given in Appendix \ref{app:errors}.


For the FGK stars in Group 2 (Fig.\ \ref{fig:boxplots2}, left panels) the systematic uncertainty (inter-quartile range) remains within a factor 2, corresponding to $1 \sigma$ in the EMD$^{\rm (s)}_{Z=1}$ case and $\leq 2 \sigma$ in the EMD$^{\rm (s)}_{Fe}$ case. 
For the M-type stars (Fig.\ \ref{fig:boxplots2}, right panels), the inter-quartile ranges are similar or even smaller, but there is a bias in the former case (EMD$^{\rm (s)}_{Z=1}$), where the X-ray fluxes computed from the synthetic EMDs tend to be higher than those derived from the spectral fits (median $R \simeq 2$). This bias is reduced in the latter case (EMD$^{\rm (s)}_{Fe}$).
We attribute this difference to contributions to the X-ray flux coming from the high-temperature tails in the EMDs of the M-type stars, that are not properly reproduced by the simple multi-temperature models employed for computing the $F_{\rm x,SF25}$ values. In fact, this bias is visible also in the reference EMD case, where the original EMDs and related abundances were employed.

Fig.\ \ref{fig:boxplots1} shows the corresponding results for Group 1 stars. Here we note that the synthetic EMDs for the FGK stars yield systematically lower values than the $F_{\rm x, SF25}$ from the best-fit models (median $R \approx 0.5$), while the opposite holds for M-type stars (median $R \approx 3$). 

Group 1 is mainly composed by stars with relatively low X-ray surface fluxes, hence the mismatch between observed (best-fit) and predicted (synthetic EMDs) values can be attributed to larger systematic uncertainties in the modeling of the EMDs for such low-activity stars.
For FGK stars, the polynomials that describe the $EM_{\rm h}(T)$ vs.\ $F_{\rm x}$ at high temperatures ($\log T > 5.5$) are constrained on the low-activity side by the EMDs reconstructed for $\alpha$~Cen\,A/B ($F_{\rm x} = 10^{3.9}$ -- $10^{4.8}$\,erg\,s$^{-1}$\,cm$^{-2}$) and for Procyon ($F_{\rm x} = 10^{4.7}$\,erg\,s$^{-1}$\,cm$^{-2}$).
For the M-type stars, the polynomials are guided by the results obtained for the Barnard's star ($F_{\rm x,SF25} = 10^{4.3}$\,erg\,s$^{-1}$\,cm$^{-2}$) and GJ 411 ($F_{\rm x,SF25} = 10^{5.1}$\,erg\,s$^{-1}$\,cm$^{-2}$), while three out of six stars in Group 1 have lower X-ray surface fluxes. Hence the synthetic EMDs for the lowest activity stars in Group 1 are extrapolated at high temperatures.
We conjecture that the mismatch can be due to an underestimation of the predicted $EM_{\rm h}(T)$ values
in corona for the FGK stars, and an overestimation for the M-type stars. However, we cannot exclude that this outcome is due to biases on the measured X-ray fluxes introduced after adoption of simple multi-component isothermal models.

A good agreement ($R < 2$) results from the comparison of the EUV fluxes for both the FGK stars and the M-type stars in Group 1.
This is explained by the better constraints available for the $EM_{\rm h}(T)$ shape at chromospheric and transition region temperatures ($\log T < 5.5$) even for stars with low $F_{\rm x}$.

\section{Summary and conclusions}
\label{sec:ending}

This work is aimed at providing a simple parameterization of the plasma emission measure distribution vs.\ temperature, from chromosphere to corona, as a function of the broad-band stellar surface X-ray flux, a proxy of the magnetic activity level in late-type stars. 
Synthetic XUV spectra derived from diverse distributions can be employed in studies of the irradiation of stellar disks or exo-planetary atmospheres, and the consequent photochemistry and photoevaporation effects.
To this aim, we employed a database of plasma emission measure distributions vs.\ temperature, reconstructed from published analyses of high-resolution FUV and X-ray line emission spectra, and the results of spectral fits of medium-resolution (CCD) spectra with simple multi-component isothermal models. 
The best-fitting models provide the broad-band X-ray fluxes that we employ as control parameter, and also a measure of the plasma metallicity or the iron abundance, that is required for the subsequent spectral synthesis.

We found that the differential emission measure at any given temperature in the range $10^4$--$10^{7.5}$\,K can be described as a polynomial function of the surface X-ray flux, but separate descriptions are required for FGK stars and M-type stars, respectively.
This parameterization allowed us to build a grid of synthetic EMDs for each spectral-type subgroup, that can be easily interpolated in order to compute predicted XUV spectra. 
This is especially important for the many late-type stars too faint for high-resolution spectroscopy.
We provide a \href{https://zenodo.org/records/17407361}{Javascript web tool} that facilitates this interpolation, and that we share  in the Zenodo archive of THE StellaR PAth project.

We verified the validity and limitations of this novel approach by comparison of the original and synthetic EMDs, and also by comparison of broad-band X-ray and EUV fluxes.
We tested different assumptions for the plasma chemical abundances, and we found that a single measurement of the iron abundance from best-fit multi-component isothermal models can be employed to predict X-ray and EUV fluxes. 
In most cases they remain within a factor 2 from the original values (based on high-resolution spectra).
Relatively larger systematic uncertainties on the predicted X-ray fluxes, within a factor 5 in the worst case (Sect.\ \ref{app:errors}), resulted only for the M-type stars with the lowest surface X-ray fluxes in our sample, i.e.\ for very low-activity low-mass stars.

We stress that the statistical uncertainties on the synthetic EMDs at coronal temperatures ($T \ge 10^6$\,K) for low-activity stars are determined mainly by the error bars on the reconstructed EMDs of relatively few objects in our sample with low surface X-ray fluxes ($F_{\rm x} < 10^5$\,\fxu) but available high-resolution X-ray spectra.
These uncertainties resulted relatively larger for the FGK stars than for the M-type stars. However, we actually have little information on a possible spread of the coronal EMDs for low-activity stars, especially M-type stars with activity levels similar to or lower than the Barnard's star or GJ 411.

Nonetheless, our results represent a significant improvement with respect to the pioneering work by \cite{Wood+2018}, based on Chandra data, because we extended the prediction of the EMDs down to chromospheric temperatures by exploiting high-resolution FUV spectra taken with HST. 
Moreover, our sample included several more M-type stars, that allowed us to find significant differences in corona between solar-type and low-mass stars.
Our approach is also complementary with respect to the long-term observation programs carried on in the framework of the MUSCLES and Mega-MUSCLE Treasury Survey \citep{France+2016, Behr+2023}.
Following the suggestion by\cite{Duvvuri+2021}, we believe to have already gathered a sufficiently comprehensive data library that makes possible to predict EMDs for stars 
too faint for FUV or X-ray high-resolution spectroscopy.

Overall, our results are significantly more promising than earlier attempts to guess X-ray and EUV irradiation levels of planets in extra-solar systems.
In particular, \cite{France+2022} suggested that different reconstructions of the EUV spectrum of Proxima Cen could imply flux discrepancies by factors up to 100.
Waiting for new much-needed space-born instrumentation for spectroscopy in the EUV domain, we propose to apply our novel approach for predicting the XUV irradiation levels in the many extra-solar systems that are currently observed with JWST. 
In particular, this approach will be employed in a further paper in preparation, focused on end-to-end simulations of irradiated planetary atmospheres that are currently included in the Reference Sample of the future Ariel mission \citep{Edards+2019}.

Coming back to the limitations of our proposed approach, we emphasize the need to improve our EMD database especially for low-activity solar-like stars and X-ray faint low-mass stars. 
In fact, the validity of our results is restricted to the range of surface X-ray fluxes in our current stellar sample. 
For stars near the low boundary of this range, the uncertainties remain large, and we discourageextrapolations.
Moreover, a long-term monitoring of a few selected targets is required to characterize the variability of the EMDs associated to magnetic-activity phenomena, such as flares and magnetic dynamo cycles.
In this respect, coordinated observations with current space instrumentation for X-ray and FUV high-resolution spectroscopy of low-activity stars should be considered as a useful investment of telescope time.

\begin{acknowledgments}
AMa and IPi acknowledge financial contribution from the INAF grant 2023 for data analysis. GMi and AMa  acknowledge also support from the European Union - Next Generation EU through 
the grant n. 2022J7ZFRA - Exo-planetary Cloudy Atmospheres and Stellar High energy 
(Exo-CASH), funded by MUR--PRIN 2022, and from the ASI--INAF agreement 2021-5-HH.2-2024.
JSF acknowledges financial support from the Agencia Estatal
de Investigaci\'on (AEI/10.13039/501100011033) of the Ministerio de Ciencia, Innovaci\'on y Universidades and the ERDF ``A way of making Europe'', through project
PID2022-137241NB-C42.
Based on observations obtained with \xmm, an ESA science mission with instruments and contributions directly funded by ESA Member States and NASA.
\end{acknowledgments}


\vspace{5mm}


\software{PINTofALE
(\citealt{PINTofALE}, \url{hea-www.harvard.edu/PINTofALE/)});
%
ATOMDB, pyatomdb \citep{Foster+2020}.
The \dataset[EMD interpolator]{\doi{10.5281/zenodo.17407361}} Javascript/HTML GUI is available on Zenodo under a GNU General Public License v3.0.
}


\bibliographystyle{aasjournal}



\appendix
\section{Stellar sample and data base}
\label{app:sample}

Here we report the main characteristics of the stellar sample employed in the present work, and the original reconstructed EMDs. In Table \ref{emduv} we gathered the targets with only high-resolution FUV spectra available (Group 1), while in Table \ref{emdxuv} are listed the stars with both FUV and X-ray high-resolution spectra (Group 2) and the data for $\alpha$ Cen A/B.

\begin{table*}[t]
\caption{\label{emduv} Characteristics and emission measure distribution vs.\ temperature for the stars in Group 1, having FUV only spectra \citep{SF25}.}
\resizebox{1.0\textwidth}{!}{
\begin{tabular}{l|cccccccccccccccc}\hline\hline
Name	 & 	GJ 357	 & 	GJ 436	 & 	GJ 486	 & 	GJ 1214	 & 	GJ 3470	 & 	GJ 9827	 & 	HD 73583	 & 	HD 97658	 & 	HD 149026	 & 	HD 189733	 & 	HD 209458	 & 	Trappist 1	 & 	$\upsilon$ And	 & 	Wasp 13	 & 	WASP 77A	 & 	TOI 836  \\
Sp.Type & M2.5V & M2.5V & M3.5V & M4.5V & M1.5V & K6V & K4V & K1V & G0V & K2V & G0V & M8.0V & F8V & G1IV-V & G8V & K7V \\
$R_*$ (R$\odot$)	 & 	0.337	 & 	0.464	 & 	0.328	 & 	0.216	 & 	0.480	 & 	0.651	 & 	0.710	 & 	0.730	 & 	1.497	 & 	0.805	 & 	1.203	 & 	0.119	 & 	1.631	 & 	1.559	 & 	0.955	 & 	0.665  \\
$L_\mathrm{X}$ (\lxu)$^a$ & 	25.66	 & 	26.14	 & 	25.50	 & 	25.88	 & 	27.53	 & 	26.53	 & 	27.82	 & 	26.92	 & 	27.40	 & 	28.36	 & 	26.84	 & 	26.06	 & 	27.83	 & 	28.43	 & 	28.13	 & 	27.41  \\
$F_\mathrm{X}$ (\fxu) & 	 3.82	 & 	 4.02	 & 	 3.68	 & 	 4.43	 & 	 5.38	 & 	 4.12	 & 	 5.33	 & 	 4.41	 & 	 4.26	 & 	 5.76	 & 	 3.89	 & 	 5.12	 & 	 4.62	 & 	 5.26	 & 	 5.38	 & 	 4.98  \\
$L_\mathrm{EUV}$ (\lxu)$^a$ & 	26.36	 & 	26.79	 & 	26.27	 & 	26.27	 & 	27.84	 & 	27.14	 & 	27.98	 & 	27.39	 & 	28.10	 & 	28.81	 & 	28.45	 & 	26.29	 & 	28.56	 & 	28.94	 & 	28.48	 & 	28.08  \\
$F_\mathrm{EUV}$ (\fxu) & 	 4.52	 & 	 4.67	 & 	 4.45	 & 	 4.82	 & 	 5.69	 & 	 4.73	 & 	 5.49	 & 	 4.88	 & 	 4.96	 & 	 6.21	 & 	 5.50	 & 	 5.35	 & 	 5.35	 & 	 5.77	 & 	 5.73	 & 	 5.65  \\
$\mathrm{[Fe/H]^b}$	 & 	-0.10	 & 	-0.29	 & 	-0.15	 & 	+0.39	 & 	 +0.20	 & 	-0.28	 & 	-0.66	 & 	-0.23	 & 	+0.36	 & 	-0.51	 & 	-0.20	 & 	+0.04	 & 	+0.24	 & 	 0.00	 & 	 0.00	 & 	-0.30  \\ \hline
$\log T$ [K] & \multicolumn{16}{c}{-- $\log EM_{\rm v}(T)$ [cm$^{-3}$] -- } \\ \hline
4.0	 & 	48.15	 & 	48.65	 & 	48.00	 & 	47.80	 & 	49.90	 & 	48.75	 & 	49.80	 & 	49.70	 & 	50.20	 & 	51.01	 & 	49.70	 & 	47.20	 & 	50.85	 & 	50.10	 & 	50.50	 & 	50.25  \\
4.1	 & 	48.05	 & 	48.55	 & 	47.90	 & 	47.60	 & 	49.70	 & 	48.60	 & 	49.60	 & 	49.55	 & 	50.10	 & 	50.86	 & 	49.60	 & 	47.10	 & 	50.70	 & 	49.90	 & 	50.40	 & 	50.10  \\
4.2	 & 	47.90	 & 	48.40	 & 	47.80	 & 	47.40	 & 	49.50	 & 	48.40	 & 	49.40	 & 	49.40	 & 	50.00	 & 	50.71	 & 	49.50	 & 	47.00	 & 	50.50	 & 	49.70	 & 	50.30	 & 	49.90  \\
4.3	 & 	47.70	 & 	48.20	 & 	47.70	 & 	47.20	 & 	49.30	 & 	48.20	 & 	49.20	 & 	49.10	 & 	49.85	 & 	50.51	 & 	49.40	 & 	46.90	 & 	50.25	 & 	49.50	 & 	50.15	 & 	49.70  \\
4.4	 & 	47.45	 & 	47.95	 & 	47.55	 & 	47.05	 & 	49.00	 & 	48.00	 & 	49.00	 & 	48.65	 & 	49.70	 & 	50.41	 & 	49.35	 & 	46.80	 & 	50.10	 & 	49.35	 & 	50.00	 & 	49.60  \\
4.5	 & 	47.10	 & 	47.50	 & 	47.35	 & 	46.90	 & 	48.65	 & 	47.80	 & 	48.80	 & 	48.30	 & 	49.50	 & 	50.16	 & 	49.25	 & 	46.60	 & 	49.85	 & 	49.10	 & 	49.90	 & 	49.55  \\
4.6	 & 	46.95	 & 	47.25	 & 	47.05	 & 	46.70	 & 	48.35	 & 	47.55	 & 	48.65	 & 	48.15	 & 	49.20	 & 	50.01	 & 	49.10	 & 	46.40	 & 	49.70	 & 	48.85	 & 	49.70	 & 	49.45  \\
4.7	 & 	46.85	 & 	47.20	 & 	46.90	 & 	46.50	 & 	48.10	 & 	47.30	 & 	48.45	 & 	48.05	 & 	48.80	 & 	49.81	 & 	48.85	 & 	46.05	 & 	49.60	 & 	48.80	 & 	49.40	 & 	49.10  \\
4.8	 & 	46.90	 & 	47.15	 & 	46.80	 & 	46.40	 & 	48.10	 & 	47.20	 & 	48.30	 & 	48.00	 & 	48.55	 & 	49.81	 & 	48.55	 & 	45.85	 & 	49.45	 & 	48.85	 & 	49.00	 & 	48.50  \\
4.9	 & 	46.85	 & 	47.20	 & 	46.75	 & 	46.45	 & 	48.35	 & 	47.25	 & 	48.20	 & 	48.25	 & 	48.50	 & 	49.76	 & 	48.60	 & 	45.95	 & 	49.40	 & 	48.90	 & 	48.90	 & 	48.40  \\
5.0	 & 	46.75	 & 	47.20	 & 	46.80	 & 	46.65	 & 	48.50	 & 	47.30	 & 	48.20	 & 	48.10	 & 	48.85	 & 	49.51	 & 	48.55	 & 	46.20	 & 	49.35	 & 	48.85	 & 	49.20	 & 	48.45  \\
5.1	 & 	46.70	 & 	47.15	 & 	46.75	 & 	46.95	 & 	48.40	 & 	47.50	 & 	48.25	 & 	47.75	 & 	48.80	 & 	49.11	 & 	48.50	 & 	46.45	 & 	48.85	 & 	48.75	 & 	48.95	 & 	48.50  \\
5.2	 & 	46.75	 & 	47.20	 & 	46.70	 & 	47.00	 & 	48.20	 & 	47.70	 & 	48.35	 & 	47.55	 & 	48.40	 & 	48.81	 & 	48.20	 & 	46.40	 & 	48.60	 & 	48.50	 & 	48.60	 & 	48.40  \\
5.3	 & 	46.65	 & 	47.15	 & 	46.60	 & 	46.90	 & 	48.15	 & 	47.65	 & 	48.40	 & 	47.60	 & 	48.20	 & 	48.61	 & 	47.80	 & 	46.30	 & 	48.60	 & 	48.35	 & 	48.50	 & 	48.30  \\
5.4	 & 	46.50	 & 	47.00	 & 	46.50	 & 	46.65	 & 	48.00	 & 	47.55	 & 	48.25	 & 	47.95	 & 	48.10	 & 	48.31	 & 	47.65	 & 	46.20	 & 	48.95	 & 	48.20	 & 	48.40	 & 	48.10  \\
5.5	 & 	46.30	 & 	46.50	 & 	46.40	 & 	46.50	 & 	47.80	 & 	47.40	 & 	48.00	 & 	47.90	 & 	48.00	 & 	48.21	 & 	47.40	 & 	46.00	 & 	48.75	 & 	48.00	 & 	48.30	 & 	47.95  \\
5.6	 & 	46.25	 & 	46.20	 & 	46.30	 & 	46.40	 & 	47.60	 & 	47.10	 & 	47.80	 & 	47.45	 & 	47.90	 & 	48.16	 & 	47.25	 & 	45.80	 & 	48.30	 & 	47.85	 & 	48.20	 & 	47.80  \\
5.7	 & 	46.20	 & 	45.90	 & 	46.20	 & 	46.20	 & 	47.50	 & 	46.75	 & 	47.65	 & 	47.15	 & 	47.80	 & 	48.26	 & 	47.20	 & 	45.70	 & 	48.05	 & 	47.70	 & 	48.10	 & 	47.90  \\
5.8	 & 	46.10	 & 	45.70	 & 	46.00	 & 	45.80	 & 	47.70	 & 	46.60	 & 	47.60	 & 	47.00	 & 	47.70	 & 	48.51	 & 	47.00	 & 	45.60	 & 	47.80	 & 	47.85	 & 	48.00	 & 	48.10  \\
5.9	 & 	46.20	 & 	45.80	 & 	46.00	 & 	46.00	 & 	48.00	 & 	46.55	 & 	47.55	 & 	47.10	 & 	47.80	 & 	48.81	 & 	47.10	 & 	45.70	 & 	47.90	 & 	48.00	 & 	47.80	 & 	48.30  \\\hline
\end{tabular}
}
\flushleft
\footnotesize
NOTES -- $^a$ X-ray (5--100\,\AA) and EUV (100-920 \AA) luminosities (SF25, Table C.3). $^b$ Iron abundance in corona (\href{https://zenodo.org/records/14500302}{SF25, Table D.1}).
\end{table*}

\begin{table*}[t]
\caption{\label{emdxuv} Characteristics and emission measure distributions vs.\ temperature for stars in Group 2 and for $\alpha$ Cen AB.}
\resizebox{1.0\textwidth}{!}{
\begin{tabular}{l|ccccccccccccccc} \hline\hline
Name	 & 	Proxima Cen	 & 	AU Mic	 & 	AD Leo	 & 	GJ 411	 & 	$\tau$ Boo	 & 	Gl 674	 & 	$\iota$ Hor	 & 	V1298 Tau	 & 	$\epsilon$ Eri	 & 	Procyon	 & 	Barnard's & \multicolumn{2}{c}{$\alpha$ Cen A} & \multicolumn{2}{c}{$\alpha$ Cen B} \\
& & & & & & & & & & & & (low) & (high) & (low) & (high) \\
Sp.Type & M5.5V & M1.5V & M3.0V & M1.5V & F7V & M2.5V & G0V & K1 & K2V & F5IV-V & M3.5V & \multicolumn{2}{c}{G2V} & \multicolumn{2}{c}{K1V}\\
$R_*/R_\odot$	 & 	0.141	 & 	0.750	 & 	0.410	 & 	0.393	 & 	1.331	 & 	0.326	 & 	1.850	 & 	1.305	 & 	0.895	 & 	2.060	 & 	0.185    & \multicolumn{2}{c}{1.22} & \multicolumn{2}{c}{0.86}\\
$L_\mathrm{X}^a$	(\lxu) & 	27.25	 & 	29.44	 & 	28.61	 & 	27.10	 & 	29.00	 & 	27.67	 & 	28.91	 & 	30.24	 & 	28.35	 & 	28.14	 & 	25.61    &  26.90    &    27.06    &  27.06    &  27.48\\
$F_\mathrm{X}$	(\fxu) & 	 6.17	 & 	 6.90	 & 	 6.60	 & 	 5.13	 & 	 5.97	 & 	 5.86	 & 	 5.59	 & 	 7.22	 & 	 5.66	 & 	 4.74	 & 	 4.29    &   3.94    &   4.10    &      4.41    &   4.83\\
$L_\mathrm{EUV}^a$ (\lxu)	 & 	27.30	 & 	29.39	 & 	28.77	 & 	26.93	 & 	29.23	 & 	28.02	 & 	29.14	 & 	30.20	 & 	28.59	 & 	29.23	 & 	26.13    &  ---      &    ----      &  ----     &  ---\\
$F_\mathrm{EUV}$	(\fxu) & 	 6.22	 & 	 6.85	 & 	 6.76	 & 	 4.96	 & 	 6.20	 & 	 6.21	 & 	 5.82	 & 	 7.18	 & 	 5.90	 & 	 5.86	 & 	 4.81    &  ---      &    ----      &  ----     &  ---\\
$\mathrm{[Fe/H]^b}$	 & 	-0.50	 & 	-0.80	 & 	-0.60	 & 	-0.50	 & 	-0.30	 & 	-0.60	 & 	-0.03	 & 	-0.89	 & 	-0.30	 & 	-0.18	 & 	 0.00    & \multicolumn{2}{c}{+0.09} & \multicolumn{2}{c}{-0.34}\\
Reference            &   1       &         1       &   1       &   1       &         1       &   1       &   2       &         3       &   4       &   5       &   1       & \multicolumn{2}{c}{6} & \multicolumn{2}{c}{6}\\
\hline
$\log T$/K  & \multicolumn{11}{c}{-- $\log EM_{\rm v}(T)$ [cm$^{-3}$] --} \\ \hline
4.0	 & 	48.57	 & 	50.70	 & 	50.50	 & 	47.70	 & 	50.70	 & 	49.75	 & 	51.14	 & 	51.99	 & 	50.60	 & 	51.58	 & 	47.20   &  --- &  --- &  --- &  ---\\
4.1	 & 	48.32	 & 	50.50	 & 	50.30	 & 	47.60	 & 	50.60	 & 	49.60	 & 	51.04	 & 	51.74	 & 	50.40	 & 	51.48	 & 	47.10   &  --- &  --- &  --- &  ---\\
4.2	 & 	48.07	 & 	50.40	 & 	50.10	 & 	47.50	 & 	50.50	 & 	49.35	 & 	50.84	 & 	51.49	 & 	50.20	 & 	51.28	 & 	47.00   &  --- &  --- &  --- &  ---\\
4.3	 & 	47.92	 & 	50.20	 & 	49.80	 & 	47.40	 & 	50.40	 & 	49.10	 & 	50.64	 & 	51.34	 & 	50.05	 & 	50.98	 & 	46.80   &  --- &  --- &  --- &  ---\\
4.4	 & 	47.82	 & 	50.05	 & 	49.45	 & 	47.30	 & 	50.30	 & 	48.95	 & 	50.44	 & 	51.24	 & 	49.90	 & 	50.88	 & 	46.60   &  --- &  --- &  --- &  ---\\
4.5	 & 	47.77	 & 	49.80	 & 	49.00	 & 	47.10	 & 	50.20	 & 	48.75	 & 	50.24	 & 	51.19	 & 	49.80	 & 	50.78	 & 	46.60   &  --- &  --- &  --- &  ---\\
4.6	 & 	47.77	 & 	49.50	 & 	48.75	 & 	46.90	 & 	50.10	 & 	48.70	 & 	49.94	 & 	51.19	 & 	49.65	 & 	50.73	 & 	46.65   &  --- &  --- &  --- &  ---\\
4.7	 & 	47.82	 & 	49.40	 & 	48.60	 & 	46.95	 & 	50.00	 & 	48.60	 & 	49.84	 & 	51.14	 & 	49.45	 & 	50.28	 & 	46.80   &  --- &  --- &  --- &  ---\\
4.8	 & 	47.77	 & 	49.35	 & 	48.70	 & 	46.95	 & 	49.95	 & 	48.75	 & 	49.74	 & 	50.99	 & 	49.20	 & 	50.08	 & 	46.90   &  --- &  --- &  --- &  ---\\
4.9	 & 	47.57	 & 	49.40	 & 	49.30	 & 	46.80	 & 	49.85	 & 	48.95	 & 	49.69	 & 	50.84	 & 	49.00	 & 	49.98	 & 	46.40   &  --- &  --- &  --- &  ---\\
5.0	 & 	47.37	 & 	49.55	 & 	49.35	 & 	46.70	 & 	49.80	 & 	48.50	 & 	49.74	 & 	50.89	 & 	49.00	 & 	49.83	 & 	46.50   &  --- &  --- &  --- &  ---\\
5.1	 & 	47.17	 & 	49.55	 & 	49.00	 & 	46.60	 & 	49.85	 & 	48.35	 & 	49.74	 & 	50.64	 & 	49.00	 & 	49.78	 & 	46.55   &  --- &  --- &  --- &  ---\\
5.2	 & 	46.97	 & 	49.40	 & 	48.65	 & 	46.65	 & 	49.80	 & 	48.30	 & 	49.44	 & 	50.54	 & 	48.90	 & 	49.73	 & 	46.30   &  --- &  --- &  --- &  ---\\
5.3	 & 	46.82	 & 	49.20	 & 	48.50	 & 	46.70	 & 	49.55	 & 	48.10	 & 	49.24	 & 	50.39	 & 	48.60	 & 	49.78	 & 	46.25   &  --- &  --- &  --- &  ---\\
5.4	 & 	46.72	 & 	49.10	 & 	48.35	 & 	46.50	 & 	49.10	 & 	47.70	 & 	48.89	 & 	50.14	 & 	48.30	 & 	49.58	 & 	46.20   &  --- &  --- &  --- &  ---\\
5.5	 & 	46.62	 & 	48.90	 & 	48.25	 & 	46.35	 & 	49.00	 & 	47.65	 & 	48.44	 & 	50.04	 & 	47.80	 & 	49.28	 & 	46.15   &  47.06 &  47.39 &  47.35 &  47.30\\
5.6	 & 	46.57	 & 	48.70	 & 	48.20	 & 	46.40	 & 	48.90	 & 	47.75	 & 	48.24	 & 	49.99	 & 	47.50	 & 	49.08	 & 	46.10   &  47.22 &  47.51 &  47.47 &  47.39\\
5.7	 & 	46.57	 & 	48.80	 & 	48.25	 & 	46.50	 & 	48.90	 & 	47.90	 & 	48.24	 & 	50.04	 & 	47.60	 & 	48.78	 & 	46.20   &  47.59 &  47.44 &  47.63 &  47.76\\
5.8	 & 	46.77	 & 	49.10	 & 	48.30	 & 	46.70	 & 	49.00	 & 	48.00	 & 	48.34	 & 	50.14	 & 	47.80	 & 	48.88	 & 	46.40   &  47.82 &  47.74 &  48.03 &  48.10\\
5.9	 & 	47.17	 & 	49.35	 & 	48.40	 & 	47.00	 & 	49.10	 & 	48.20	 & 	48.44	 & 	50.39	 & 	48.20	 & 	49.28	 & 	46.60   &  48.07 &  48.07 &  48.37 &  48.45\\
6.0	 & 	47.72	 & 	49.70	 & 	48.60	 & 	47.30	 & 	49.30	 & 	48.70	 & 	48.74	 & 	50.84	 & 	48.65	 & 	49.73	 & 	46.90   &  48.67 &  48.44 &  48.69 &  48.68\\
6.1	 & 	48.27	 & 	50.10	 & 	49.00	 & 	47.70	 & 	49.55	 & 	49.50	 & 	48.94	 & 	51.54	 & 	49.30	 & 	49.98	 & 	47.30   &  48.63 &  48.78 &  48.75 &  48.78\\
6.2	 & 	48.77	 & 	50.40	 & 	50.05	 & 	48.00	 & 	50.05	 & 	49.40	 & 	49.49	 & 	51.84	 & 	49.85	 & 	50.18	 & 	47.70   &  47.99 &  48.46 &  48.94 &  49.33\\
6.3	 & 	48.52	 & 	50.70	 & 	50.35	 & 	48.15	 & 	50.35	 & 	48.80	 & 	49.84	 & 	51.39	 & 	50.15	 & 	50.28	 & 	47.80   &  48.37 &  48.76 &  49.21 &  49.64\\
6.4	 & 	48.32	 & 	50.40	 & 	49.65	 & 	47.90	 & 	50.00	 & 	48.65	 & 	49.54	 & 	51.49	 & 	50.00	 & 	50.08	 & 	47.20   &  48.25 &  48.54 &  48.78 &  49.54\\
6.5	 & 	48.42	 & 	50.75	 & 	49.95	 & 	48.20	 & 	50.30	 & 	48.60	 & 	49.64	 & 	51.59	 & 	50.00	 & 	49.48	 & 	47.10   &  47.08 &  47.67 &  48.44 &  49.43\\
6.6	 & 	48.77	 & 	51.00	 & 	50.20	 & 	48.40	 & 	51.15	 & 	48.70	 & 	49.74	 & 	51.69	 & 	50.20	 & 	48.58	 & 	47.20   &  45.83 &  46.51 &  47.79 &  48.48\\
6.7	 & 	49.22	 & 	51.30	 & 	50.40	 & 	48.50	 & 	51.30	 & 	48.90	 & 	50.39	 & 	51.89	 & 	50.30	 & 	47.88	 & 	47.30   &  45.21 &  45.96 &  47.56 &  48.33\\
6.8	 & 	49.42	 & 	51.50	 & 	50.60	 & 	48.60	 & 	50.40	 & 	49.10	 & 	50.69	 & 	52.24	 & 	50.30	 & 	47.48	 & 	47.50   &  44.59 &  45.42 &  47.80 &  48.67\\
6.9	 & 	49.12	 & 	51.80	 & 	50.85	 & 	48.50	 & 	49.80	 & 	49.50	 & 	49.94	 & 	52.54	 & 	50.15	 & 	47.18	 & 	47.95   &  43.98 &  44.88 &  47.40 &  47.95\\
7.0	 & 	49.07	 & 	51.60	 & 	50.80	 & 	48.70	 & 	49.50	 & 	50.00	 & 	50.04	 & 	52.24	 & 	49.40	 & 	46.88	 & 	48.15   &  43.69 &  44.76 &  46.57 &  47.69\\
7.1	 & 	49.27	 & 	51.80	 & 	50.55	 & 	48.80	 & 	49.30	 & 	49.80	 & 	49.34	 & 	52.39	 & 	48.80	 & 	46.68	 & 	47.70   &  43.41 &  44.64 &  46.18 &  47.11\\
7.2	 & 	48.87	 & 	51.20	 & 	50.40	 & 	48.50	 & 	49.00	 & 	49.30	 & 	48.69	 & 	52.54	 & 	48.30	 & 	46.38	 & 	47.10   &  43.13 &  44.52 &  45.92 &  46.16\\
7.3	 & 	48.47	 & 	50.70	 & 	50.20	 & 	48.10	 & 	48.70	 & 	48.80	 & 	47.74	 & 	52.39	 & 	47.70	 & 	45.98	 & 	46.50   &  42.84 &  44.40 &  45.43 &  45.57\\
7.4	 & 	47.97	 & 	50.20	 & 	49.80	 & 	47.30	 & 	48.30	 & 	48.10	 & 	47.54	 & 	51.99	 & 	47.50	 & 	45.58	 & 	46.00   &  42.56 &  44.28 &  44.94 &  44.97\\ \hline
\end{tabular}
}
\flushleft
\footnotesize
NOTES -- $^a$ X-ray (5--100\,\AA) and EUV (100-920 \AA) luminosities. $^b$ Iron abundance in corona.

\noindent
References -- (1) \cite{SF25}, (2) HR 810, \cite{SanzForcada+2019}, (3) \cite{Maggio+2023}, (4) \cite{Chadney+2015}, (5) \cite{SanzForcada+2003,SanzForcada+2004}, (6) \cite{Wood+2018}. 
\end{table*}

\section{How-to guide and software tools}
\label{app:recipe}

\begin{table*}[]
    \caption{Synthetic EMD grid for FGK stars.}
    \label{tab:fgksynthgrid}
    \resizebox{\textwidth}{!}{
\begin{tabular}{l|ccccccccc} \hline \hline 
$\log F_{\rm x,ref}$ & 	3.5	 & 	4.0	 & 	4.5	 & 	5.0	 & 	5.5	 & 	6.0	 & 	6.5	 & 	7.0	 & 	7.5  \\ \hline
$\log T$ [K] & \multicolumn{9}{c}{-- $\log EM_{\rm h}(T)$ [cm$^{-5}$ K$^{-1}$] -- } \\ \hline
4.0	 & 	27.11 ( 26.69 $-$ 27.53 )	 & 	27.97 ( 27.77 $-$ 28.17 )	 & 	28.62 ( 28.49 $-$ 28.75 )	 & 	28.93 ( 28.81 $-$ 29.05 )	 & 	29.05 ( 28.94 $-$ 29.16 )	 & 	29.16 ( 28.99 $-$ 29.34 )	 & 	29.59 ( 29.37 $-$ 29.82 )	 & 	30.03 ( 29.75 $-$ 30.30 )	 & 	30.46 ( 30.13 $-$ 30.79 )  \\
4.1	 & 	27.07 ( 26.61 $-$ 27.53 )	 & 	27.91 ( 27.69 $-$ 28.13 )	 & 	28.54 ( 28.41 $-$ 28.68 )	 & 	28.84 ( 28.71 $-$ 28.97 )	 & 	28.95 ( 28.84 $-$ 29.07 )	 & 	29.05 ( 28.88 $-$ 29.23 )	 & 	29.47 ( 29.24 $-$ 29.70 )	 & 	29.88 ( 29.60 $-$ 30.16 )	 & 	30.29 ( 29.97 $-$ 30.62 )  \\
4.2	 & 	26.98 ( 26.49 $-$ 27.48 )	 & 	27.76 ( 27.53 $-$ 28.00 )	 & 	28.36 ( 28.22 $-$ 28.50 )	 & 	28.66 ( 28.53 $-$ 28.79 )	 & 	28.79 ( 28.67 $-$ 28.91 )	 & 	28.91 ( 28.73 $-$ 29.09 )	 & 	29.29 ( 29.06 $-$ 29.53 )	 & 	29.68 ( 29.39 $-$ 29.97 )	 & 	30.06 ( 29.72 $-$ 30.41 )  \\
4.3	 & 	26.92 ( 26.42 $-$ 27.42 )	 & 	27.59 ( 27.35 $-$ 27.82 )	 & 	28.11 ( 27.97 $-$ 28.25 )	 & 	28.42 ( 28.29 $-$ 28.55 )	 & 	28.60 ( 28.48 $-$ 28.72 )	 & 	28.77 ( 28.58 $-$ 28.95 )	 & 	29.13 ( 28.87 $-$ 29.38 )	 & 	29.48 ( 29.16 $-$ 29.81 )	 & 	29.84 ( 29.45 $-$ 30.24 )  \\
4.4	 & 	26.82 ( 26.32 $-$ 27.32 )	 & 	27.37 ( 27.14 $-$ 27.61 )	 & 	27.84 ( 27.71 $-$ 27.98 )	 & 	28.17 ( 28.04 $-$ 28.31 )	 & 	28.42 ( 28.31 $-$ 28.54 )	 & 	28.67 ( 28.48 $-$ 28.85 )	 & 	29.02 ( 28.74 $-$ 29.30 )	 & 	29.38 ( 29.01 $-$ 29.75 )	 & 	29.73 ( 29.27 $-$ 30.19 )  \\
4.5	 & 	26.73 ( 26.28 $-$ 27.19 )	 & 	27.21 ( 26.99 $-$ 27.42 )	 & 	27.63 ( 27.49 $-$ 27.77 )	 & 	27.97 ( 27.84 $-$ 28.10 )	 & 	28.27 ( 28.15 $-$ 28.38 )	 & 	28.56 ( 28.38 $-$ 28.75 )	 & 	28.94 ( 28.66 $-$ 29.22 )	 & 	29.32 ( 28.95 $-$ 29.69 )	 & 	29.70 ( 29.23 $-$ 30.16 )  \\
4.6	 & 	26.49 ( 26.01 $-$ 26.97 )	 & 	26.99 ( 26.76 $-$ 27.22 )	 & 	27.43 ( 27.29 $-$ 27.57 )	 & 	27.78 ( 27.65 $-$ 27.91 )	 & 	28.08 ( 27.96 $-$ 28.19 )	 & 	28.39 ( 28.21 $-$ 28.58 )	 & 	28.84 ( 28.55 $-$ 29.13 )	 & 	29.28 ( 28.90 $-$ 29.67 )	 & 	29.73 ( 29.24 $-$ 30.21 )  \\
4.7	 & 	26.18 ( 25.66 $-$ 26.71 )	 & 	26.73 ( 26.49 $-$ 26.97 )	 & 	27.21 ( 27.07 $-$ 27.35 )	 & 	27.58 ( 27.44 $-$ 27.71 )	 & 	27.89 ( 27.77 $-$ 28.01 )	 & 	28.23 ( 28.04 $-$ 28.42 )	 & 	28.73 ( 28.44 $-$ 29.01 )	 & 	29.23 ( 28.85 $-$ 29.61 )	 & 	29.73 ( 29.25 $-$ 30.20 )  \\
4.8	 & 	26.12 ( 25.62 $-$ 26.62 )	 & 	26.62 ( 26.39 $-$ 26.85 )	 & 	27.06 ( 26.93 $-$ 27.19 )	 & 	27.42 ( 27.29 $-$ 27.55 )	 & 	27.75 ( 27.64 $-$ 27.87 )	 & 	28.11 ( 27.92 $-$ 28.30 )	 & 	28.62 ( 28.33 $-$ 28.91 )	 & 	29.13 ( 28.75 $-$ 29.51 )	 & 	29.64 ( 29.16 $-$ 30.12 )  \\
4.9	 & 	26.19 ( 25.70 $-$ 26.68 )	 & 	26.61 ( 26.38 $-$ 26.83 )	 & 	26.99 ( 26.86 $-$ 27.13 )	 & 	27.34 ( 27.21 $-$ 27.47 )	 & 	27.68 ( 27.56 $-$ 27.80 )	 & 	28.05 ( 27.86 $-$ 28.24 )	 & 	28.54 ( 28.26 $-$ 28.82 )	 & 	29.03 ( 28.66 $-$ 29.40 )	 & 	29.52 ( 29.05 $-$ 29.98 )  \\
5.0	 & 	26.51 ( 26.06 $-$ 26.96 )	 & 	26.74 ( 26.52 $-$ 26.95 )	 & 	26.99 ( 26.85 $-$ 27.12 )	 & 	27.28 ( 27.15 $-$ 27.41 )	 & 	27.62 ( 27.50 $-$ 27.74 )	 & 	28.00 ( 27.81 $-$ 28.19 )	 & 	28.45 ( 28.20 $-$ 28.70 )	 & 	28.90 ( 28.59 $-$ 29.21 )	 & 	29.35 ( 28.98 $-$ 29.73 )  \\
5.1	 & 	26.55 ( 26.09 $-$ 27.00 )	 & 	26.73 ( 26.51 $-$ 26.95 )	 & 	26.94 ( 26.81 $-$ 27.08 )	 & 	27.22 ( 27.09 $-$ 27.35 )	 & 	27.55 ( 27.43 $-$ 27.66 )	 & 	27.93 ( 27.75 $-$ 28.11 )	 & 	28.37 ( 28.10 $-$ 28.64 )	 & 	28.81 ( 28.44 $-$ 29.17 )	 & 	29.24 ( 28.79 $-$ 29.70 )  \\
5.2	 & 	26.28 ( 25.81 $-$ 26.74 )	 & 	26.56 ( 26.34 $-$ 26.78 )	 & 	26.84 ( 26.70 $-$ 26.97 )	 & 	27.12 ( 27.00 $-$ 27.25 )	 & 	27.43 ( 27.32 $-$ 27.55 )	 & 	27.78 ( 27.60 $-$ 27.97 )	 & 	28.23 ( 27.94 $-$ 28.51 )	 & 	28.67 ( 28.28 $-$ 29.06 )	 & 	29.12 ( 28.63 $-$ 29.61 )  \\
5.3	 & 	25.83 ( 25.38 $-$ 26.27 )	 & 	26.34 ( 26.13 $-$ 26.55 )	 & 	26.76 ( 26.62 $-$ 26.89 )	 & 	27.03 ( 26.90 $-$ 27.16 )	 & 	27.25 ( 27.13 $-$ 27.36 )	 & 	27.50 ( 27.31 $-$ 27.68 )	 & 	27.98 ( 27.69 $-$ 28.27 )	 & 	28.46 ( 28.06 $-$ 28.86 )	 & 	28.94 ( 28.43 $-$ 29.45 )  \\
5.4	 & 	25.37 ( 24.94 $-$ 25.80 )	 & 	26.14 ( 25.94 $-$ 26.34 )	 & 	26.69 ( 26.55 $-$ 26.83 )	 & 	26.93 ( 26.79 $-$ 27.06 )	 & 	27.02 ( 26.90 $-$ 27.14 )	 & 	27.16 ( 26.98 $-$ 27.34 )	 & 	27.71 ( 27.45 $-$ 27.98 )	 & 	28.27 ( 27.93 $-$ 28.61 )	 & 	28.83 ( 28.40 $-$ 29.25 )  \\
5.5	 & 	25.08 ( 24.62 $-$ 25.54 )	 & 	25.97 ( 25.77 $-$ 26.17 )	 & 	26.55 ( 26.41 $-$ 26.69 )	 & 	26.75 ( 26.62 $-$ 26.89 )	 & 	26.77 ( 26.65 $-$ 26.89 )	 & 	26.85 ( 26.67 $-$ 27.03 )	 & 	27.47 ( 27.24 $-$ 27.70 )	 & 	28.09 ( 27.81 $-$ 28.36 )	 & 	28.71 ( 28.38 $-$ 29.03 )  \\
5.6	 & 	25.04 ( 24.56 $-$ 25.52 )	 & 	25.83 ( 25.63 $-$ 26.02 )	 & 	26.33 ( 26.19 $-$ 26.47 )	 & 	26.52 ( 26.39 $-$ 26.66 )	 & 	26.58 ( 26.46 $-$ 26.69 )	 & 	26.71 ( 26.53 $-$ 26.89 )	 & 	27.36 ( 27.13 $-$ 27.59 )	 & 	28.01 ( 27.74 $-$ 28.29 )	 & 	28.66 ( 28.34 $-$ 28.98 )  \\
5.7	 & 	25.01 ( 24.53 $-$ 25.48 )	 & 	25.68 ( 25.50 $-$ 25.87 )	 & 	26.13 ( 26.00 $-$ 26.26 )	 & 	26.37 ( 26.25 $-$ 26.50 )	 & 	26.53 ( 26.41 $-$ 26.64 )	 & 	26.76 ( 26.58 $-$ 26.94 )	 & 	27.41 ( 27.19 $-$ 27.64 )	 & 	28.06 ( 27.79 $-$ 28.33 )	 & 	28.71 ( 28.40 $-$ 29.03 )  \\
5.8	 & 	25.03 ( 24.59 $-$ 25.47 )	 & 	25.70 ( 25.52 $-$ 25.87 )	 & 	26.17 ( 26.05 $-$ 26.28 )	 & 	26.43 ( 26.32 $-$ 26.55 )	 & 	26.62 ( 26.51 $-$ 26.73 )	 & 	26.88 ( 26.70 $-$ 27.05 )	 & 	27.53 ( 27.31 $-$ 27.76 )	 & 	28.19 ( 27.91 $-$ 28.47 )	 & 	28.85 ( 28.52 $-$ 29.17 )  \\
5.9	 & 	25.44 ( 25.02 $-$ 25.86 )	 & 	26.06 ( 25.88 $-$ 26.23 )	 & 	26.49 ( 26.36 $-$ 26.62 )	 & 	26.71 ( 26.58 $-$ 26.84 )	 & 	26.86 ( 26.74 $-$ 26.98 )	 & 	27.09 ( 26.91 $-$ 27.28 )	 & 	27.77 ( 27.54 $-$ 28.00 )	 & 	28.44 ( 28.16 $-$ 28.72 )	 & 	29.11 ( 28.79 $-$ 29.44 )  \\
6.0	 & 	25.98 ( 25.56 $-$ 26.40 )	 & 	26.58 ( 26.39 $-$ 26.77 )	 & 	26.97 ( 26.82 $-$ 27.12 )	 & 	27.11 ( 26.96 $-$ 27.26 )	 & 	27.17 ( 27.02 $-$ 27.31 )	 & 	27.34 ( 27.14 $-$ 27.54 )	 & 	28.09 ( 27.84 $-$ 28.34 )	 & 	28.84 ( 28.54 $-$ 29.14 )	 & 	29.59 ( 29.24 $-$ 29.95 )  \\
6.1	 & 	26.11 ( 25.67 $-$ 26.54 )	 & 	26.90 ( 26.69 $-$ 27.11 )	 & 	27.43 ( 27.26 $-$ 27.60 )	 & 	27.57 ( 27.39 $-$ 27.75 )	 & 	27.57 ( 27.41 $-$ 27.73 )	 & 	27.69 ( 27.47 $-$ 27.90 )	 & 	28.50 ( 28.24 $-$ 28.77 )	 & 	29.32 ( 29.01 $-$ 29.64 )	 & 	30.14 ( 29.78 $-$ 30.50 )  \\
6.2	 & 	25.84 ( 25.38 $-$ 26.30 )	 & 	26.96 ( 26.73 $-$ 27.19 )	 & 	27.74 ( 27.56 $-$ 27.91 )	 & 	27.98 ( 27.79 $-$ 28.17 )	 & 	28.00 ( 27.83 $-$ 28.17 )	 & 	28.08 ( 27.85 $-$ 28.31 )	 & 	28.85 ( 28.57 $-$ 29.12 )	 & 	29.61 ( 29.30 $-$ 29.93 )	 & 	30.38 ( 30.02 $-$ 30.74 )  \\
6.3	 & 	25.29 ( 24.76 $-$ 25.81 )	 & 	26.72 ( 26.47 $-$ 26.97 )	 & 	27.75 ( 27.59 $-$ 27.92 )	 & 	28.14 ( 27.96 $-$ 28.31 )	 & 	28.24 ( 28.07 $-$ 28.40 )	 & 	28.31 ( 28.09 $-$ 28.54 )	 & 	28.99 ( 28.73 $-$ 29.26 )	 & 	29.67 ( 29.37 $-$ 29.97 )	 & 	30.35 ( 30.01 $-$ 30.68 )  \\
6.4	 & 	24.74 ( 24.09 $-$ 25.38 )	 & 	26.40 ( 26.10 $-$ 26.70 )	 & 	27.63 ( 27.45 $-$ 27.81 )	 & 	28.14 ( 27.95 $-$ 28.32 )	 & 	28.30 ( 28.14 $-$ 28.47 )	 & 	28.39 ( 28.16 $-$ 28.62 )	 & 	28.98 ( 28.72 $-$ 29.23 )	 & 	29.57 ( 29.29 $-$ 29.84 )	 & 	30.15 ( 29.86 $-$ 30.45 )  \\
6.5	 & 	24.46 ( 23.50 $-$ 25.43 )	 & 	25.91 ( 25.45 $-$ 26.36 )	 & 	27.08 ( 26.86 $-$ 27.30 )	 & 	27.78 ( 27.57 $-$ 27.98 )	 & 	28.24 ( 28.06 $-$ 28.41 )	 & 	28.58 ( 28.35 $-$ 28.82 )	 & 	29.08 ( 28.82 $-$ 29.34 )	 & 	29.58 ( 29.30 $-$ 29.87 )	 & 	30.08 ( 29.77 $-$ 30.40 )  \\
6.6	 & 	23.59 ( 22.24 $-$ 24.95 )	 & 	24.99 ( 24.34 $-$ 25.64 )	 & 	26.28 ( 26.01 $-$ 26.56 )	 & 	27.33 ( 27.10 $-$ 27.57 )	 & 	28.24 ( 28.05 $-$ 28.42 )	 & 	28.95 ( 28.72 $-$ 29.18 )	 & 	29.38 ( 29.12 $-$ 29.64 )	 & 	29.82 ( 29.53 $-$ 30.10 )	 & 	30.25 ( 29.94 $-$ 30.57 )  \\
6.7	 & 	22.29 ( 20.58 $-$ 24.00 )	 & 	23.95 ( 23.11 $-$ 24.78 )	 & 	25.55 ( 25.22 $-$ 25.89 )	 & 	27.00 ( 26.74 $-$ 27.26 )	 & 	28.29 ( 28.10 $-$ 28.48 )	 & 	29.30 ( 29.06 $-$ 29.53 )	 & 	29.67 ( 29.40 $-$ 29.93 )	 & 	30.04 ( 29.75 $-$ 30.33 )	 & 	30.41 ( 30.10 $-$ 30.72 )  \\
6.8	 & 	20.47 ( 18.27 $-$ 22.66 )	 & 	22.77 ( 21.69 $-$ 23.86 )	 & 	24.91 ( 24.52 $-$ 25.30 )	 & 	26.66 ( 26.39 $-$ 26.93 )	 & 	28.14 ( 27.94 $-$ 28.33 )	 & 	29.25 ( 29.02 $-$ 29.48 )	 & 	29.75 ( 29.49 $-$ 30.01 )	 & 	30.24 ( 29.96 $-$ 30.53 )	 & 	30.74 ( 30.43 $-$ 31.06 )  \\
6.9	 & 	18.80 ( 16.40 $-$ 21.20 )	 & 	21.88 ( 20.70 $-$ 23.05 )	 & 	24.54 ( 24.14 $-$ 24.93 )	 & 	26.39 ( 26.13 $-$ 26.65 )	 & 	27.79 ( 27.59 $-$ 27.98 )	 & 	28.80 ( 28.56 $-$ 29.03 )	 & 	29.54 ( 29.28 $-$ 29.81 )	 & 	30.29 ( 30.00 $-$ 30.58 )	 & 	31.04 ( 30.72 $-$ 31.35 )  \\
7.0	 & 	18.25 ( 15.60 $-$ 20.90 )	 & 	21.45 ( 20.16 $-$ 22.75 )	 & 	24.17 ( 23.76 $-$ 24.58 )	 & 	25.99 ( 25.73 $-$ 26.24 )	 & 	27.32 ( 27.13 $-$ 27.52 )	 & 	28.32 ( 28.08 $-$ 28.56 )	 & 	29.29 ( 29.02 $-$ 29.55 )	 & 	30.25 ( 29.96 $-$ 30.54 )	 & 	31.22 ( 30.90 $-$ 31.53 )  \\
7.1	 & 	18.99 ( 16.20 $-$ 21.78 )	 & 	21.62 ( 20.26 $-$ 22.97 )	 & 	23.89 ( 23.48 $-$ 24.31 )	 & 	25.54 ( 25.29 $-$ 25.78 )	 & 	26.85 ( 26.66 $-$ 27.04 )	 & 	27.94 ( 27.70 $-$ 28.18 )	 & 	29.06 ( 28.80 $-$ 29.32 )	 & 	30.18 ( 29.90 $-$ 30.47 )	 & 	31.31 ( 30.99 $-$ 31.62 )  \\
7.2	 & 	20.40 ( 17.53 $-$ 23.27 )	 & 	22.13 ( 20.73 $-$ 23.53 )	 & 	23.73 ( 23.30 $-$ 24.16 )	 & 	25.09 ( 24.84 $-$ 25.34 )	 & 	26.34 ( 26.15 $-$ 26.53 )	 & 	27.54 ( 27.30 $-$ 27.77 )	 & 	28.82 ( 28.55 $-$ 29.08 )	 & 	30.10 ( 29.81 $-$ 30.39 )	 & 	31.38 ( 31.07 $-$ 31.70 )  \\
7.3	 & 	20.83 ( 17.54 $-$ 24.11 )	 & 	22.10 ( 20.49 $-$ 23.71 )	 & 	23.35 ( 22.87 $-$ 23.83 )	 & 	24.59 ( 24.33 $-$ 24.84 )	 & 	25.83 ( 25.63 $-$ 26.02 )	 & 	27.11 ( 26.87 $-$ 27.36 )	 & 	28.56 ( 28.29 $-$ 28.82 )	 & 	30.00 ( 29.71 $-$ 30.29 )	 & 	31.44 ( 31.13 $-$ 31.75 )  \\
7.4	 & 	21.09 ( 17.57 $-$ 24.60 )	 & 	22.05 ( 20.33 $-$ 23.77 )	 & 	23.08 ( 22.57 $-$ 23.58 )	 & 	24.24 ( 23.98 $-$ 24.49 )	 & 	25.49 ( 25.29 $-$ 25.68 )	 & 	26.83 ( 26.58 $-$ 27.08 )	 & 	28.31 ( 28.04 $-$ 28.58 )	 & 	29.79 ( 29.50 $-$ 30.08 )	 & 	31.27 ( 30.95 $-$ 31.58 )  \\ \hline
\end{tabular}
}
\end{table*}

\begin{table*}[]
    \caption{Synthetic EMD grid for M stars.}
    \label{tab:msynthgrid}
    \resizebox{\textwidth}{!}{
\begin{tabular}{l|ccccccccc} \hline\hline 
$\log F_{\rm x,ref}$	 & 	3.5	 & 	4.0	 & 	4.5	 & 	5.0	 & 	5.5	 & 	6.0	 & 	6.5	 & 	7.0	 & 	7.5  \\ \hline
$\log T$ [K] & \multicolumn{9}{c}{-- $\log EM_{\rm h}(T)$ [cm$^{-5}$ K$^{-1}$] -- } \\ \hline
4.0	 & 	28.01 ( 27.68 $-$ 28.33 )	 & 	27.39 ( 27.24 $-$ 27.53 )	 & 	27.36 ( 27.20 $-$ 27.52 )	 & 	27.76 ( 27.62 $-$ 27.90 )	 & 	28.34 ( 28.20 $-$ 28.49 )	 & 	28.98 ( 28.81 $-$ 29.15 )	 & 	29.41 ( 29.23 $-$ 29.58 )	 & 	29.51 ( 29.21 $-$ 29.81 )	 & 	29.54 ( 29.09 $-$ 30.00 )  \\
4.1	 & 	27.94 ( 27.59 $-$ 28.29 )	 & 	27.31 ( 27.15 $-$ 27.46 )	 & 	27.26 ( 27.09 $-$ 27.43 )	 & 	27.63 ( 27.48 $-$ 27.78 )	 & 	28.19 ( 28.03 $-$ 28.34 )	 & 	28.81 ( 28.63 $-$ 28.99 )	 & 	29.25 ( 29.06 $-$ 29.44 )	 & 	29.38 ( 29.05 $-$ 29.71 )	 & 	29.45 ( 28.95 $-$ 29.96 )  \\
4.2	 & 	27.81 ( 27.45 $-$ 28.17 )	 & 	27.18 ( 27.02 $-$ 27.34 )	 & 	27.12 ( 26.94 $-$ 27.30 )	 & 	27.47 ( 27.31 $-$ 27.63 )	 & 	28.01 ( 27.85 $-$ 28.17 )	 & 	28.62 ( 28.42 $-$ 28.81 )	 & 	29.05 ( 28.84 $-$ 29.26 )	 & 	29.19 ( 28.82 $-$ 29.56 )	 & 	29.28 ( 28.71 $-$ 29.84 )  \\
4.3	 & 	27.61 ( 27.25 $-$ 27.97 )	 & 	27.01 ( 26.84 $-$ 27.17 )	 & 	26.95 ( 26.76 $-$ 27.13 )	 & 	27.29 ( 27.12 $-$ 27.45 )	 & 	27.80 ( 27.64 $-$ 27.97 )	 & 	28.39 ( 28.20 $-$ 28.59 )	 & 	28.83 ( 28.61 $-$ 29.04 )	 & 	28.99 ( 28.61 $-$ 29.37 )	 & 	29.10 ( 28.52 $-$ 29.68 )  \\
4.4	 & 	27.37 ( 27.03 $-$ 27.71 )	 & 	26.82 ( 26.66 $-$ 26.98 )	 & 	26.78 ( 26.60 $-$ 26.96 )	 & 	27.12 ( 26.96 $-$ 27.28 )	 & 	27.63 ( 27.47 $-$ 27.79 )	 & 	28.21 ( 28.01 $-$ 28.40 )	 & 	28.64 ( 28.43 $-$ 28.84 )	 & 	28.80 ( 28.43 $-$ 29.18 )	 & 	28.92 ( 28.34 $-$ 29.50 )  \\
4.5	 & 	27.14 ( 26.81 $-$ 27.47 )	 & 	26.59 ( 26.44 $-$ 26.75 )	 & 	26.58 ( 26.41 $-$ 26.76 )	 & 	26.95 ( 26.80 $-$ 27.11 )	 & 	27.48 ( 27.33 $-$ 27.64 )	 & 	28.06 ( 27.87 $-$ 28.25 )	 & 	28.44 ( 28.24 $-$ 28.64 )	 & 	28.52 ( 28.16 $-$ 28.88 )	 & 	28.55 ( 27.99 $-$ 29.10 )  \\
4.6	 & 	26.91 ( 26.58 $-$ 27.24 )	 & 	26.41 ( 26.26 $-$ 26.57 )	 & 	26.42 ( 26.24 $-$ 26.59 )	 & 	26.78 ( 26.62 $-$ 26.94 )	 & 	27.29 ( 27.13 $-$ 27.45 )	 & 	27.85 ( 27.66 $-$ 28.04 )	 & 	28.22 ( 28.02 $-$ 28.42 )	 & 	28.31 ( 27.95 $-$ 28.66 )	 & 	28.34 ( 27.79 $-$ 28.89 )  \\
4.7	 & 	26.76 ( 26.42 $-$ 27.10 )	 & 	26.30 ( 26.15 $-$ 26.45 )	 & 	26.32 ( 26.15 $-$ 26.49 )	 & 	26.69 ( 26.54 $-$ 26.85 )	 & 	27.20 ( 27.04 $-$ 27.37 )	 & 	27.75 ( 27.55 $-$ 27.94 )	 & 	28.11 ( 27.90 $-$ 28.31 )	 & 	28.18 ( 27.81 $-$ 28.55 )	 & 	28.20 ( 27.62 $-$ 28.77 )  \\
4.8	 & 	26.71 ( 26.37 $-$ 27.05 )	 & 	26.23 ( 26.07 $-$ 26.38 )	 & 	26.25 ( 26.08 $-$ 26.43 )	 & 	26.64 ( 26.49 $-$ 26.80 )	 & 	27.18 ( 27.02 $-$ 27.34 )	 & 	27.75 ( 27.56 $-$ 27.95 )	 & 	28.13 ( 27.93 $-$ 28.33 )	 & 	28.21 ( 27.86 $-$ 28.57 )	 & 	28.24 ( 27.68 $-$ 28.79 )  \\
4.9	 & 	26.71 ( 26.36 $-$ 27.07 )	 & 	26.21 ( 26.05 $-$ 26.37 )	 & 	26.23 ( 26.05 $-$ 26.40 )	 & 	26.62 ( 26.47 $-$ 26.78 )	 & 	27.18 ( 27.02 $-$ 27.35 )	 & 	27.79 ( 27.59 $-$ 27.99 )	 & 	28.23 ( 28.03 $-$ 28.44 )	 & 	28.39 ( 28.02 $-$ 28.76 )	 & 	28.50 ( 27.91 $-$ 29.08 )  \\
5.0	 & 	26.63 ( 26.27 $-$ 26.99 )	 & 	26.21 ( 26.05 $-$ 26.37 )	 & 	26.26 ( 26.08 $-$ 26.44 )	 & 	26.65 ( 26.49 $-$ 26.81 )	 & 	27.20 ( 27.03 $-$ 27.36 )	 & 	27.80 ( 27.61 $-$ 28.00 )	 & 	28.28 ( 28.07 $-$ 28.48 )	 & 	28.52 ( 28.12 $-$ 28.91 )	 & 	28.71 ( 28.07 $-$ 29.34 )  \\
5.1	 & 	26.51 ( 26.13 $-$ 26.88 )	 & 	26.24 ( 26.07 $-$ 26.40 )	 & 	26.33 ( 26.14 $-$ 26.51 )	 & 	26.69 ( 26.53 $-$ 26.84 )	 & 	27.17 ( 27.01 $-$ 27.33 )	 & 	27.72 ( 27.53 $-$ 27.91 )	 & 	28.18 ( 27.99 $-$ 28.38 )	 & 	28.49 ( 28.13 $-$ 28.85 )	 & 	28.76 ( 28.19 $-$ 29.33 )  \\
5.2	 & 	26.33 ( 25.94 $-$ 26.72 )	 & 	26.27 ( 26.10 $-$ 26.43 )	 & 	26.41 ( 26.22 $-$ 26.59 )	 & 	26.70 ( 26.54 $-$ 26.86 )	 & 	27.08 ( 26.92 $-$ 27.24 )	 & 	27.52 ( 27.33 $-$ 27.71 )	 & 	27.94 ( 27.74 $-$ 28.14 )	 & 	28.30 ( 27.95 $-$ 28.65 )	 & 	28.65 ( 28.11 $-$ 29.19 )  \\
5.3	 & 	26.18 ( 25.80 $-$ 26.55 )	 & 	26.19 ( 26.03 $-$ 26.36 )	 & 	26.36 ( 26.17 $-$ 26.54 )	 & 	26.63 ( 26.47 $-$ 26.79 )	 & 	26.97 ( 26.81 $-$ 27.13 )	 & 	27.35 ( 27.15 $-$ 27.54 )	 & 	27.71 ( 27.51 $-$ 27.91 )	 & 	28.04 ( 27.70 $-$ 28.39 )	 & 	28.37 ( 27.83 $-$ 28.90 )  \\
5.4	 & 	26.03 ( 25.65 $-$ 26.41 )	 & 	26.07 ( 25.91 $-$ 26.23 )	 & 	26.23 ( 26.06 $-$ 26.41 )	 & 	26.49 ( 26.33 $-$ 26.65 )	 & 	26.81 ( 26.65 $-$ 26.97 )	 & 	27.17 ( 26.97 $-$ 27.36 )	 & 	27.54 ( 27.33 $-$ 27.75 )	 & 	27.89 ( 27.51 $-$ 28.26 )	 & 	28.24 ( 27.66 $-$ 28.82 )  \\
5.5	 & 	25.84 ( 25.47 $-$ 26.21 )	 & 	25.90 ( 25.75 $-$ 26.06 )	 & 	26.08 ( 25.91 $-$ 26.25 )	 & 	26.35 ( 26.20 $-$ 26.50 )	 & 	26.67 ( 26.52 $-$ 26.83 )	 & 	27.04 ( 26.84 $-$ 27.23 )	 & 	27.41 ( 27.19 $-$ 27.62 )	 & 	27.75 ( 27.35 $-$ 28.16 )	 & 	28.10 ( 27.47 $-$ 28.73 )  \\
5.6	 & 	25.68 ( 25.31 $-$ 26.05 )	 & 	25.74 ( 25.59 $-$ 25.90 )	 & 	25.94 ( 25.77 $-$ 26.11 )	 & 	26.24 ( 26.09 $-$ 26.39 )	 & 	26.59 ( 26.44 $-$ 26.75 )	 & 	26.98 ( 26.79 $-$ 27.17 )	 & 	27.34 ( 27.13 $-$ 27.55 )	 & 	27.66 ( 27.26 $-$ 28.05 )	 & 	27.96 ( 27.35 $-$ 28.57 )  \\
5.7	 & 	25.55 ( 25.18 $-$ 25.92 )	 & 	25.60 ( 25.45 $-$ 25.76 )	 & 	25.84 ( 25.67 $-$ 26.00 )	 & 	26.20 ( 26.05 $-$ 26.35 )	 & 	26.61 ( 26.46 $-$ 26.76 )	 & 	27.04 ( 26.86 $-$ 27.22 )	 & 	27.40 ( 27.20 $-$ 27.59 )	 & 	27.64 ( 27.28 $-$ 28.01 )	 & 	27.87 ( 27.30 $-$ 28.44 )  \\
5.8	 & 	25.46 ( 25.12 $-$ 25.79 )	 & 	25.56 ( 25.41 $-$ 25.70 )	 & 	25.84 ( 25.68 $-$ 26.01 )	 & 	26.26 ( 26.11 $-$ 26.40 )	 & 	26.72 ( 26.57 $-$ 26.86 )	 & 	27.18 ( 27.01 $-$ 27.35 )	 & 	27.55 ( 27.38 $-$ 27.72 )	 & 	27.80 ( 27.48 $-$ 28.12 )	 & 	28.02 ( 27.52 $-$ 28.53 )  \\
5.9	 & 	25.61 ( 25.28 $-$ 25.93 )	 & 	25.80 ( 25.62 $-$ 25.99 )	 & 	26.12 ( 25.93 $-$ 26.31 )	 & 	26.52 ( 26.36 $-$ 26.68 )	 & 	26.96 ( 26.80 $-$ 27.12 )	 & 	27.41 ( 27.24 $-$ 27.58 )	 & 	27.80 ( 27.63 $-$ 27.97 )	 & 	28.12 ( 27.82 $-$ 28.41 )	 & 	28.42 ( 27.97 $-$ 28.87 )  \\
6.0	 & 	26.01 ( 25.67 $-$ 26.35 )	 & 	26.29 ( 26.04 $-$ 26.54 )	 & 	26.62 ( 26.40 $-$ 26.85 )	 & 	27.00 ( 26.81 $-$ 27.19 )	 & 	27.39 ( 27.21 $-$ 27.57 )	 & 	27.78 ( 27.60 $-$ 27.97 )	 & 	28.15 ( 27.98 $-$ 28.33 )	 & 	28.49 ( 28.20 $-$ 28.77 )	 & 	28.81 ( 28.39 $-$ 29.23 )  \\
6.1	 & 	26.50 ( 26.09 $-$ 26.91 )	 & 	26.85 ( 26.51 $-$ 27.20 )	 & 	27.21 ( 26.92 $-$ 27.49 )	 & 	27.55 ( 27.33 $-$ 27.78 )	 & 	27.90 ( 27.69 $-$ 28.12 )	 & 	28.25 ( 28.06 $-$ 28.45 )	 & 	28.60 ( 28.42 $-$ 28.78 )	 & 	28.94 ( 28.64 $-$ 29.24 )	 & 	29.28 ( 28.84 $-$ 29.73 )  \\
6.2	 & 	27.04 ( 26.60 $-$ 27.47 )	 & 	27.34 ( 26.95 $-$ 27.73 )	 & 	27.62 ( 27.30 $-$ 27.93 )	 & 	27.88 ( 27.64 $-$ 28.13 )	 & 	28.20 ( 27.98 $-$ 28.42 )	 & 	28.55 ( 28.35 $-$ 28.75 )	 & 	28.97 ( 28.79 $-$ 29.15 )	 & 	29.45 ( 29.16 $-$ 29.75 )	 & 	29.94 ( 29.50 $-$ 30.38 )  \\
6.3	 & 	27.10 ( 26.68 $-$ 27.52 )	 & 	27.35 ( 26.99 $-$ 27.71 )	 & 	27.59 ( 27.29 $-$ 27.89 )	 & 	27.83 ( 27.60 $-$ 28.07 )	 & 	28.18 ( 27.97 $-$ 28.39 )	 & 	28.58 ( 28.39 $-$ 28.77 )	 & 	29.09 ( 28.92 $-$ 29.26 )	 & 	29.67 ( 29.38 $-$ 29.96 )	 & 	30.26 ( 29.82 $-$ 30.70 )  \\
6.4	 & 	27.03 ( 26.63 $-$ 27.43 )	 & 	27.23 ( 26.88 $-$ 27.57 )	 & 	27.42 ( 27.14 $-$ 27.70 )	 & 	27.62 ( 27.40 $-$ 27.84 )	 & 	27.97 ( 27.77 $-$ 28.17 )	 & 	28.42 ( 28.24 $-$ 28.60 )	 & 	29.03 ( 28.86 $-$ 29.21 )	 & 	29.76 ( 29.47 $-$ 30.06 )	 & 	30.51 ( 30.06 $-$ 30.95 )  \\
6.5	 & 	26.83 ( 26.48 $-$ 27.18 )	 & 	27.04 ( 26.74 $-$ 27.35 )	 & 	27.26 ( 27.00 $-$ 27.52 )	 & 	27.47 ( 27.26 $-$ 27.69 )	 & 	27.86 ( 27.66 $-$ 28.05 )	 & 	28.34 ( 28.15 $-$ 28.52 )	 & 	28.99 ( 28.81 $-$ 29.18 )	 & 	29.77 ( 29.45 $-$ 30.09 )	 & 	30.56 ( 30.08 $-$ 31.05 )  \\
6.6	 & 	26.55 ( 26.18 $-$ 26.93 )	 & 	26.89 ( 26.56 $-$ 27.21 )	 & 	27.22 ( 26.95 $-$ 27.49 )	 & 	27.56 ( 27.34 $-$ 27.77 )	 & 	28.01 ( 27.81 $-$ 28.21 )	 & 	28.54 ( 28.36 $-$ 28.72 )	 & 	29.19 ( 28.99 $-$ 29.39 )	 & 	29.93 ( 29.57 $-$ 30.28 )	 & 	30.68 ( 30.14 $-$ 31.22 )  \\
6.7	 & 	26.70 ( 26.32 $-$ 27.07 )	 & 	27.05 ( 26.73 $-$ 27.37 )	 & 	27.40 ( 27.14 $-$ 27.67 )	 & 	27.76 ( 27.54 $-$ 27.97 )	 & 	28.25 ( 28.05 $-$ 28.44 )	 & 	28.81 ( 28.63 $-$ 29.00 )	 & 	29.53 ( 29.33 $-$ 29.72 )	 & 	30.33 ( 29.98 $-$ 30.68 )	 & 	31.16 ( 30.63 $-$ 31.69 )  \\
6.8	 & 	26.96 ( 26.58 $-$ 27.33 )	 & 	27.28 ( 26.96 $-$ 27.60 )	 & 	27.61 ( 27.34 $-$ 27.88 )	 & 	27.94 ( 27.72 $-$ 28.15 )	 & 	28.42 ( 28.23 $-$ 28.62 )	 & 	29.00 ( 28.82 $-$ 29.18 )	 & 	29.74 ( 29.56 $-$ 29.92 )	 & 	30.60 ( 30.28 $-$ 30.91 )	 & 	31.47 ( 31.00 $-$ 31.94 )  \\
6.9	 & 	27.29 ( 26.91 $-$ 27.67 )	 & 	27.57 ( 27.25 $-$ 27.89 )	 & 	27.85 ( 27.58 $-$ 28.11 )	 & 	28.12 ( 27.91 $-$ 28.33 )	 & 	28.57 ( 28.38 $-$ 28.77 )	 & 	29.13 ( 28.95 $-$ 29.31 )	 & 	29.86 ( 29.70 $-$ 30.03 )	 & 	30.73 ( 30.46 $-$ 31.00 )	 & 	31.61 ( 31.20 $-$ 32.01 )  \\
7.0	 & 	27.45 ( 27.07 $-$ 27.83 )	 & 	27.71 ( 27.39 $-$ 28.03 )	 & 	27.98 ( 27.71 $-$ 28.24 )	 & 	28.24 ( 28.03 $-$ 28.45 )	 & 	28.67 ( 28.48 $-$ 28.87 )	 & 	29.20 ( 29.02 $-$ 29.38 )	 & 	29.91 ( 29.75 $-$ 30.07 )	 & 	30.73 ( 30.47 $-$ 31.00 )	 & 	31.58 ( 31.18 $-$ 31.98 )  \\
7.1	 & 	27.16 ( 26.78 $-$ 27.53 )	 & 	27.49 ( 27.17 $-$ 27.81 )	 & 	27.83 ( 27.56 $-$ 28.10 )	 & 	28.17 ( 27.95 $-$ 28.38 )	 & 	28.62 ( 28.43 $-$ 28.82 )	 & 	29.15 ( 28.98 $-$ 29.33 )	 & 	29.81 ( 29.65 $-$ 29.97 )	 & 	30.55 ( 30.28 $-$ 30.83 )	 & 	31.31 ( 30.90 $-$ 31.73 )  \\
7.2	 & 	26.36 ( 25.99 $-$ 26.74 )	 & 	26.87 ( 26.54 $-$ 27.19 )	 & 	27.37 ( 27.10 $-$ 27.64 )	 & 	27.87 ( 27.66 $-$ 28.09 )	 & 	28.41 ( 28.22 $-$ 28.60 )	 & 	28.97 ( 28.79 $-$ 29.14 )	 & 	29.56 ( 29.39 $-$ 29.72 )	 & 	30.17 ( 29.89 $-$ 30.45 )	 & 	30.79 ( 30.37 $-$ 31.21 )  \\
7.3	 & 	25.61 ( 25.23 $-$ 25.99 )	 & 	26.18 ( 25.85 $-$ 26.50 )	 & 	26.74 ( 26.48 $-$ 27.01 )	 & 	27.31 ( 27.10 $-$ 27.52 )	 & 	27.90 ( 27.71 $-$ 28.09 )	 & 	28.51 ( 28.33 $-$ 28.69 )	 & 	29.15 ( 28.98 $-$ 29.31 )	 & 	29.81 ( 29.53 $-$ 30.08 )	 & 	30.47 ( 30.05 $-$ 30.88 )  \\
7.4	 & 	25.17 ( 24.78 $-$ 25.55 )	 & 	25.72 ( 25.40 $-$ 26.05 )	 & 	26.28 ( 26.02 $-$ 26.54 )	 & 	26.84 ( 26.63 $-$ 27.04 )	 & 	27.44 ( 27.25 $-$ 27.63 )	 & 	28.07 ( 27.90 $-$ 28.24 )	 & 	28.75 ( 28.58 $-$ 28.91 )	 & 	29.46 ( 29.18 $-$ 29.73 )	 & 	30.18 ( 29.76 $-$ 30.59 )  \\ \hline 
\end{tabular}
}
\end{table*}

We implemented a simple web interface in HTML and Javascript to perform an interpolation of the synthetic EMD grids for FGK and M-type stars (Table \ref{tab:fgksynthgrid} and Table \ref{tab:msynthgrid}). 
The script\footnote{The EMD interpolator can be found  \href{https://zenodo.org/records/17407361}{at this link}.} receives as an input the surface X-ray flux, $F_{\rm x}$, the stellar radius, and the choice of the grid based on the spectral type (FGK vs.\ M-type).
The two closest values of $F_{\rm x}$ bracketing the input value in the selected grid are identified, and a linear interpolation between the two corresponding EMDs is made. 
The same is done for the values of the lower and upper $1\sigma$ boundaries of the synthetic grids, in order to obtain the uncertainty range of the $EM_{\rm h}(T)$ values. 
The resulting interpolated EMD is plotted and tabulated, and there is the option to download it in comma-separated values format.

Furthermore, we outline here how to synthesize a spectrum using the synthetic EMD.
A Python script can read the EMD table in csv format generated from the Javascript tool described above, and then it can import the functions to use ATOMDB (package pyatomdb, \citealp{Foster2023}) for obtaining the line and continuum plasma emissivities, for each temperature bin.
The synthetic spectrum is then obtained by summing the contributions from all emissivities weighted by the chemical abundances and by the $EM(T)$ values. 
The resulting spectrum can be saved both as a CSV file and as a FITS file containing both the spectrum and the input EMD.
We implemented a web solution based on Python and {\em flask} to provide a {\em frontend} that receives the input EMD, a global metallicity value or a table with individual element abundances, and produces a graph of the resulting spectrum with options to save it in CSV or FITS format. A {\em backend} manages the ATOMDB session and calculates the spectrum. 

\section{Error budget}
\label{app:errors}

\begin{figure*}
    \centering
    \includegraphics[width=0.95\linewidth]{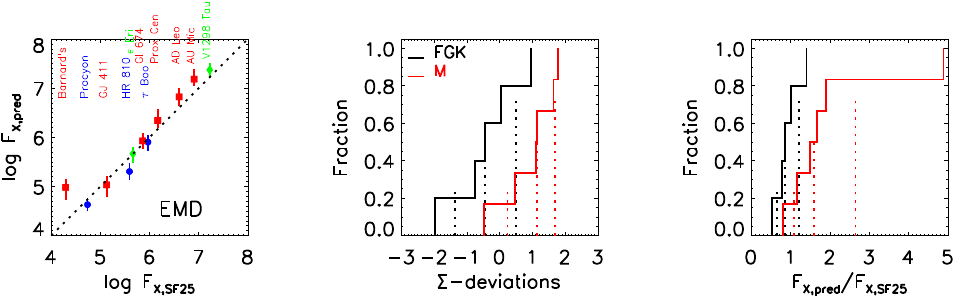}
    \includegraphics[width=0.95\linewidth]{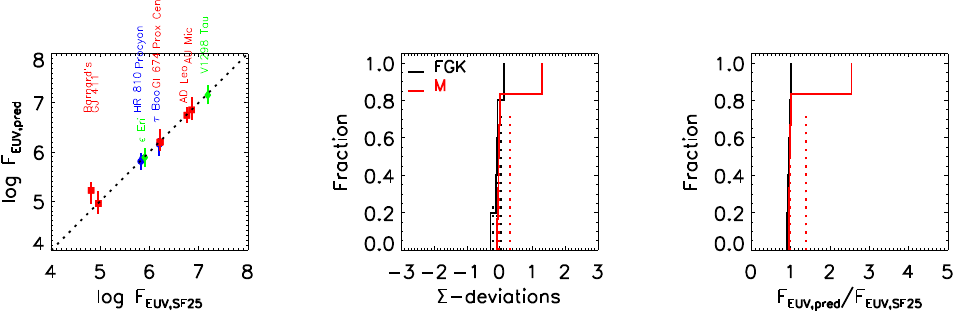}
    \caption{
    Comparison similar to Fig. \ref{fig:errors_Fe}, but with predicted X-ray and EUV fluxes based on full reconstructed EMDs and measured abundances \citep{SF25}, for stars in Group 2.
    }
    \label{fig:errors}
\end{figure*}

X-ray fluxes derived from CCD-resolution spectra may differ from the fluxes based on reconstructed EMDs based on high-resolution spectra. 
Such a comparison is possible for the stars in our Group 2 subsample\footnote{This comparison cannot be performed for the stars in Group 1, because full EMDs are not available, and X-ray fluxes can be only computed from best-fit models to CCD spectra}, and we show the results in Fig.\ \ref{fig:errors} (upper panels).
This is due to the basic difference between a full EMD and its approximation with few isothermal components. Moreover, multi-component isothermal models usually include only the iron abundance among the best-fit parameters, and the abundances of few more elements in case of spectra with high S/N ratios. 

The central 50\% range of the $\Sigma$ distributions (leftmost boxplot in Fig.\ \ref{fig:boxplots2}, labelled "EMD" case) results $-1.37$ -- $+0.49 \sigma$ for FGK stars, and $+0.24$ -- $+1.68 \sigma$ for M-type stars in Group 2. The corresponding ranges of the $R$ ratios are 0.7--1.2 for FGK stars and 1.1--2.7 for M-type stars.
We note that the distributions for the M-type stars are shifted systematically toward larger values with respect to the stars with earlier spectral types. 
We attribute this difference to contributions to the X-ray flux coming from the high-temperature tails in the EMDs of the M-type stars, that are not properly reproduced by the simple multi-temperature models employed for computing the $F_{\rm x, obs}$ values. 

This is a basal component of the full error budget, to be taken into account when synthetic EMDs rather than reconstructed EMDs are used for computing stellar XUV spectra and for predictions of their high-energy emission.

In a similar way, we also compared the EUV fluxes, and we obtained the following inter-quartile ranges: $-0.20$ -- $+0.04 \sigma$ for FGK stars, and $-0.05$ -- $+0.33 \sigma$ for M-type stars; 0.97--1.0 and 0.92--0.99, respectively, in terms of flux ratios, $R$. 
In this case, there is a nearly perfect agreement between the values based on the original EMDs of the stars in Group 2 \citep{SF25} and our computations, because EUV fluxes are always derived from full EMDs.

In the following, we describe quantitatively the results of the other cases introduced in Sect.\ \ref{sec:discuss}, based on the same synthetic EMDs but different assumptions about the plasma metallicity.

If all chemical abundances are fixes to solar values (EMD$^{\rm (s)}_{Z=1}$ case), the inter-quartile ranges (Figure \ref{fig:boxplots2}) of the $\Sigma$ and $R$ distributions are larger than in the previous "EMD" case, for the stars in Group 2. 
This is expected, especially for the EUV fluxes, because here we employed the synthetic EMDs. 
However, the systematic uncertainty remains within about $1 \sigma$ or a factor of two among FGK stars. 
For the M-type stars, the X-ray fluxes computed from the synthetic EMDs tend to be higher than those derived from the spectral fits (median $R \simeq 2$), in a way similar to the "EMD" case described above.

For the stars in Group 1 (Fig.\ \ref{fig:boxplots1}) we note that the synthetic EMDs for the FGK stars yield systematically lower values than the $F_{\rm x, obs}$ from the best-fit models (median $R = 0.54$), while the opposite holds for M-type stars (median $R = 3.1$). The bias remains within a factor $\sim 2$ in the earlier subgroup, and within $R \sim 5$ for the later type stars, in the worst occurrence.
We propose an explanation for this bias in the main text (Sect.\ \ref{sec:discuss}).

A better agreement (within $1 \sigma$ or a factor $R < 2$) is shown by the comparison of the EUV fluxes for both the FGK stars and the M-type stars.

In the next case (EMD$^{\rm (s)}_{\rm Fe})$
we fixed only the iron abundance.
Looking at the X-ray fluxes, the inter-quartile ranges resulted $-1.35$ -- $+0.20 \sigma$ or $R = 0.67$--1.24 for FGK stars, and $-0.10$ -- $+1.14 \sigma$ or $R = 0.95$--2.21 for M-type stars.
The corresponding ranges for the EUV fluxes are $-2.06$ -- $+0.69 \sigma$ or $R = 0.61$--1.43 for the earlier spectral type subgroup, and $-1.07$ -- $+0.80 \sigma$ or $R = 0.63$--1.53 for M-type stars.

The boxplots in Fig.\ \ref{fig:boxplots2} and Fig.\ \ref{fig:boxplots1} indicate smaller deviations and/or narrower 
inter-quartile ranges for many distributions with respect to the previous case.

The flux comparison for the stars in Group 1 remains the most critical, especially for the X-ray fluxes of the M-type stars, that tend to remain overestimated with respect to the best-fit values.
The inter-quartile range is small and just above $1 \sigma$, due to the large error bars on the X-ray fluxes, but in terms of flux ratios the range is $R = 1.65$--4.27, with a median of 3.50.
The EUV fluxes are always recovered more accurately, with $R < 2$ at the 75\% quantile level.

\section{Additional figures 
\label{app:addfigs}}

\begin{figure*}
    \centering
    \includegraphics[width=0.8\linewidth]{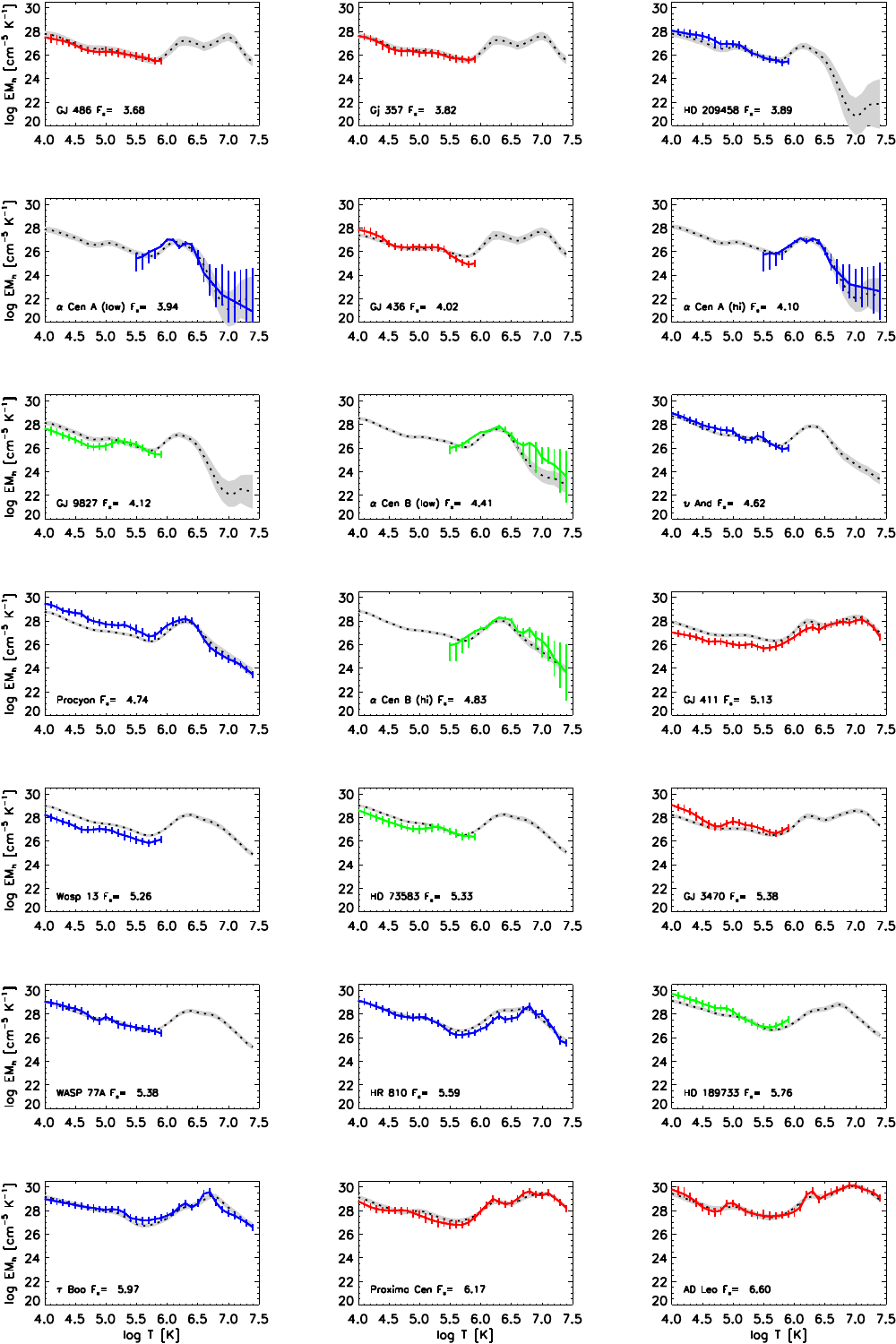}
    \caption{Reconstructed (solid colored lines) and synthetic (dotted lines) emission measure distributions for other stars in the original sample. F--G-type stars (in blue), K-type stars (green), and M-type stars (red), are all sorted by increasing surface X-ray fluxes. Dotted lines and gray-shaded $1 \sigma$ uncertainty regions were obtained by interpolating the curves in Fig.\ \ref{fig:grids}.
    }
    \label{fig:compare_more}
\end{figure*}

In Figure \ref{fig:compare_more} we show the comparison of the original and predicted EMDs for all the other stars in our sample, beyond those already selected for Figure \ref{fig:compare}. 

\end{document}